\documentclass{aa}  

\usepackage{hyperref}
\usepackage{graphicx}
\usepackage{txfonts}
\usepackage{float}
\usepackage{orcidlink}
\begin{document}

\title{An Earth-Sized Planet in a 5.4h Orbit Around a Nearby K dwarf\thanks{Based on observations obtained with the Hobby-Eberly Telescope (HET), which is a joint
project of the University of Texas at Austin, the Pennsylvania State University, Ludwig-Maximilians-Universitaet Muenchen, and Georg-August Universitaet Goettingen. The HET is
named in honor of its principal benefactors, William P. Hobby and Robert E. Eberly.}}

\subtitle{}

\author{
Kaya Han Taş\orcidlink{0009-0008-3175-8005}\inst{1}
\and
Gudmundur Stefansson\orcidlink{0000-0001-7409-5688}\inst{1}
\and
Syarief N. M. Fariz\orcidlink{0009-0008-4256-4740}\inst{1}
\and
Esha Garg\orcidlink{0009-0007-7020-7807}\inst{1}
\and
Juan I. Espinoza-Retamal\orcidlink{0000-0001-9480-8526}\inst{2,3,1}
\and
Elise Koo\orcidlink{0009-0007-0740-0954}\inst{1}
\and
David Bruijne\orcidlink{0009-0006-0218-6269}\inst{1}
\and
Jacob Luhn\orcidlink{0000-0002-4927-9925}\inst{4}
\and
Eric B. Ford\orcidlink{0000-0001-6545-639X}\inst{5,6,7,8}
\and
Suvrath Mahadevan\orcidlink{0000-0001-9596-7983}\inst{5,6}
\and
Sarah E. Logsdon\orcidlink{0000-0002-9632-9382}\inst{9}
\and
Caleb I. Cañas\orcidlink{0000-0003-4835-0619}\inst{10}
\and
Te Han\orcidlink{0000-0002-7127-7643}\inst{21}
\and
Mark E. Everett\orcidlink{0000-0002-0885-7215}\inst{9}
\and
Jaime A. Alvarado-Montes\orcidlink{0000-0003-0353-9741}\inst{11, 12}
\and
Cullen Blake\orcidlink{0000-0002-6096-1749}\inst{14}
\and
William D. Cochran\orcidlink{0000-0001-9662-3496}\inst{22}
\and
Jiayin Dong\orcidlink{0000-0002-3610-6953}\inst{15,16}
\and
Rachel B. Fernandes\orcidlink{0000-0002-3853-7327}\inst{5,6}\thanks{President's Postdoctoral Fellow}
\and
Mark R. Giovinazzi\orcidlink{0000-0002-0078-5288}\inst{17}
\and
Samuel Halverson\orcidlink{0000-0003-1312-9391}\inst{4}
\and
Shubham Kanodia\orcidlink{0000-0001-8401-4300}\inst{18}
\and
Daniel Krolikowski\orcidlink{0000-0001-9626-0613}\inst{19}
\and
Michael McElwain\orcidlink{0000-0003-0241-8956}\inst{10}
\and
Joe Ninan\orcidlink{0000-0001-8720-5612}\inst{20}
\and
Leonardo A. Paredes\orcidlink{0000-0003-1324-0495}\inst{19}
\and
Paul Robertson\orcidlink{0000-0003-0149-9678}\inst{21}
\and
Christian Schwab\orcidlink{0000-0002-4046-987X}\inst{13}
}

\institute{Anton Pannekoek Institute for Astronomy, University of Amsterdam, Science Park 904, 1098 XH Amsterdam, The Netherlands
\and
Instituto de Astrof\'isica, Pontificia Universidad Cat\'olica de Chile, Av. Vicu\~na Mackenna 4860, 782-0436 Macul, Santiago, Chile
\and
Millennium Institute for Astrophysics, Santiago, Chile
\and
Jet Propulsion Laboratory, California Institute of Technology, 4800 Oak Grove Drive, Pasadena, CA 91109, USA
\and
Department of Astronomy \& Astrophysics, 525 Davey Laboratory, The Pennsylvania State University, University Park, PA 16802, USA
\and
Center for Exoplanets and Habitable Worlds, 525 Davey Laboratory, The Pennsylvania State University, University Park, PA 16802, USA
\and
Institute for Computational and Data Sciences, The Pennsylvania State University, University Park, PA 16802, USA
\and
Center for Astrostatistics, 525 Davey Laboratory, The Pennsylvania State University, University Park, PA 16802, USA
\and
U.S. National Science Foundation National Optical-Infrared Astronomy Research Laboratory, 950 N. Cherry Avenue, Tucson, AZ 85719, USA
\and
NASA Goddard Space Flight Center, Greenbelt, MD 20771, USA
\and
Australian Astronomical Optics, Macquarie University, Balaclava Road, North Ryde, NSW 2109, Australia
\and
Astrophysics and Space Technologies Research Centre, Macquarie University, Balaclava Road, North Ryde, NSW 2109, Australia
\and
School of Mathematical and Physical Sciences, Macquarie University, Balaclava Road, North Ryde, NSW 2109, Australia
\and
Department of Physics and Astronomy, University of Pennsylvania, 209 South 33rd Street, Philadelphia, PA 19104, USA
\and
Center for Computational Astrophysics, Flatiron Institute, 162 Fifth Avenue, New York, NY 10010, USA
\and
Department of Astronomy, University of Illinois at Urbana-Champaign, Urbana, IL 61801, USA
\and
Department of Physics and Astronomy, Amherst College, Amherst, MA 01002, USA
\and
Earth and Planets Laboratory, Carnegie Institution for Science, 5241 Broad Branch Road, NW, Washington, DC 20015, USA
\and
Steward Observatory, University of Arizona, 933 N. Cherry Ave, Tucson, AZ 85721, USA
\and
Department of Astronomy and Astrophysics, Tata Institute of Fundamental Research, Homi Bhabha Road, Colaba, Mumbai 400005, India
\and
Department of Physics \& Astronomy, The University of California, Irvine, Irvine, CA 92697, USA
\and
McDonald Observatory and Center for Planetary Systems Habitability, The University of Texas at Austin, Austin, TX 78730, USA
}

\date{}
 
\abstract{We present the discovery and confirmation of the ultra-short period (USP) planet TOI-2431 b orbiting a nearby ($d\sim36$ pc) late K star ($T_{\mathrm{eff}}$ = $4109 \pm 28 \, {\rm K}$) using observations from the Transiting Exoplanet Survey Satellite (TESS), precise radial velocities with the NEID and the Habitable-zone Planet Finder (HPF) spectrographs, as well as ground-based high contrast imaging from NESSI. TOI-2431 b has a period of 5 hours and 22 minutes, making it one of the shortest-period exoplanets known to date. TOI-2431 b has a radius of $1.536 \pm 0.033\, \rm{R_\oplus}$, and a mass of $6.2 \pm 1.2\, \rm{M_\oplus}$, suggesting it has a density compatible with an Earth-like composition and, due to its high irradiation, is likely a ‘lava-world’ with a $T_{\mathrm{eq}}$ = $2063 \pm 30 \, {\rm K}$. We estimate that the current orbital period is only 30\% larger than the Roche-limit orbital period, and that it has an expected orbital decay timescale of only $\sim$31 Myr. Finally, due to the brightness of the host star ($V = 10.9$, $K = 7.6$), TOI-2431 b has a high Emission Spectroscopy Metric of 27, making it one of the best USP systems for atmospheric phase-curve analysis.}


\titlerunning{An Earth-Sized Planet in a 5.4h Orbit Around a Nearby K dwarf}
\authorrunning{Taş et al.}
\maketitle
%

% ---------------------------------------------------
% ---------------------------------------------------
% ---------------------------------------------------
\section{Introduction}
Ultra-short period (USP) planets are planets that have an orbital period smaller than one day \citep{Sahu_2006, 2016_Adams_etal, 2025_Goyal_Wang}. Of the $\sim$ 6000 exoplanets discovered so far, about 150 are confirmed USP planets as per the NASA Exoplanet Archive\footnote{\url{https://exoplanetarchive.ipac.caltech.edu/}} \citep{Akeson_2013,Christiansen2025}. Most USP planets have radii $<2 \,~\rm{R_{\oplus}}$ \citep{winn2018kepler}. They tend to have Earth-like compositions \citep{Dai2019-be}, although some are consistent with compositions enhanced in iron as compared to the Earth \citep{Price2020-ht,Uzsoy2021-bx}. Due to their close-in orbits, USP planets tend to have surface temperatures in excess of 2000 K, suggesting their surfaces are likely molten. 

Using data from the \textit{Kepler} mission, \citet{Sanchis-Ojeda2014-od} found that the occurrence rate of USP planets is $1.1\pm0.4\%$ for M dwarfs, $0.51\pm0.07\%$ for G dwarfs, and only $0.15\pm0.05\%$ for F dwarfs, suggesting that the presence of a USP planet depends on the spectral type of the star. Further, \citet{winn2018kepler} noted that USP planets are often found in compact multi-planet systems, with high orbital period ratios, where the orbital period of the innermost USP planet and its closest neighbor typically has a ratio of four or more \citep{Steffen_Farr_2013, winn2018kepler, Pu_Lai_2019}. This ratio is larger than the value generally seen in multi-planet systems discovered by \textit{Kepler} \citep{Fabrycky_2017}. Further, \cite{Dai_2018} showed that the dispersion of orbital inclinations among transiting planets tends to be larger when a USP planet is part of the system.

The origin of USP planets is not fully understood, and a few different formation scenarios have been proposed. One scenario is that rocky USP planets represent the exposed cores of hot Jupiters  \citep{Jackson_2013ApJ, Valsecchi_2015ApJ, Konigl_2017ApJ}. However, \citet{Winn_2017} found the metallicity distribution of USP-planet host stars to be different from that of hot Jupiter host stars. In addition, they found that the metallicity distributions of stars hosting rocky USP planets, and stars hosting $2-4 \,\rm{R_\oplus}$ planets with orbital periods of a few days, are identical. From this perspective, USP planets could represent the exposed cores of such smaller ($2-4\,\rm{R_\oplus}$) gaseous planets and/or super-Earths \citep[e.g.,][]{Lundkvist_2016, Lee2017-ms}. Since USP planets are typically found in the star's dust sublimation region, it is unlikely that they formed at their present location \citep{USP}, and therefore likely migrated to their current orbits. A number of USP migration mechanisms have been proposed, most involving tidal dissipation from different sources and initial conditions \citep{2010_Schlaufman, Lee2017-ms}. \cite{Petrovich} studied compact multi-planet systems and suggested the possibility of eccentricity excitation from secular dynamical chaos. \cite{Pu_Lai_2019} suggested that the outer planets tidally interact with each other and with the innermost planet, damping its eccentricity to close to zero and shrinking its semi-major axis in a quasi-equilibrium state. \cite{2020_Millholland} proposed that obliquity-driven tidal migration could be a way to obtain USP orbits. Finally, \cite{Tu_2025NatAs} found that the occurrence of USP planets increases with stellar age, and suggested that different tidal migration pathways may be responsible for younger and older USP planets. 

Due to their close-in and short-period orbits, USP planets are some of the most favorable systems for thermal phase-curve and secondary eclipse observations. Owing to their high stellar irradiation, it is likely that any hydrogen-helium (H/He) atmosphere they may have had, has been completely stripped away \citep{Sanchis-Ojeda2014-od, Lundkvist_2016, 2017_Lopez}. Their atmospheric and surface characteristics---such as albedo, phase shifts, and temperature differences between the day-and night sides---can shed light on the primary surface mineralogy and/or the presence of a secondary atmosphere \citep{2012_Hu, 2016_Demory, 2019_Kreidberg, 2022_Whittaker, 2024_Zhang, Dai2024-qd}.

The Transiting Exoplanet Survey Satellite \citep[TESS;][]{Ricker_2014} is enabling the detection of USP planets around nearby bright stars, facilitating precise mass constraints with ground-based Doppler spectroscopy, and phase-curve observations with JWST. LHS 3844 b \citep{2019_Vanderspek}, a rocky super-earth orbiting an M dwarf 15 pc away, was the first USP planet found by TESS, and more recent discoveries include GJ 367 b \citep{2021_Lam}, TOI-6255 b  \citep{Dai2024-qd}, and TOI-6324 b \citep{RenaLee_2025ApJ}.

In this paper, we present the discovery of TOI-2431 b, a USP planet with a period of $\sim$$0.224$ days, making it the sixth shortest period planet known to date. TOI-2431 is a bright ($V=10.9 \, \mathrm{mag}$, $K=7.6 \, \mathrm{mag}$) K7V star at a distance of 36 pc. TOI-2431 was identified as a TESS Objects of Interest, and statistically validated by \citet{2021ApJS..254...39G}. Here, we present a detailed characterization of the system, including precise RV follow-up observations to determine its mass and thereby confirming its planetary nature. Additionally, due to the short period of the planet, we show that TOI-2431 b is very close to its Roche limit, and has one of the highest Emission Spectroscopy Metrics \citep[ESM;][]{Kempton_2018} with JWST enabling the possibility for high signal-to-noise-ratio phase-curve observations.

This paper is structured as follows. In Section~\ref{sec:observationandanalysis}, we describe the observations and data reduction. In Section~\ref{sec:stellarparams}, we discuss the parameters of the host star. In Section~\ref{sec:analysis} we discuss the Transit and RV joint analysis, and we present our planet parameter constraints in Section~\ref{sec:results}. In Section~\ref{sec:discussion}, we discuss our search for additional planets in the system, as well as the composition, tidal decay, tidal distortion, and potential for future observations of TOI-2431 b. We conclude with a summary of our findings in Section~\ref{sec:conclusion}. 

% ---------------------------------------------------
% ---------------------------------------------------
% ---------------------------------------------------
\section{Observations and Data Reduction} \label{sec:observationandanalysis}

% ---------------------------------------------------
\subsection{TESS Photometry}
\begin{figure}
\centering
\includegraphics[width=\columnwidth]{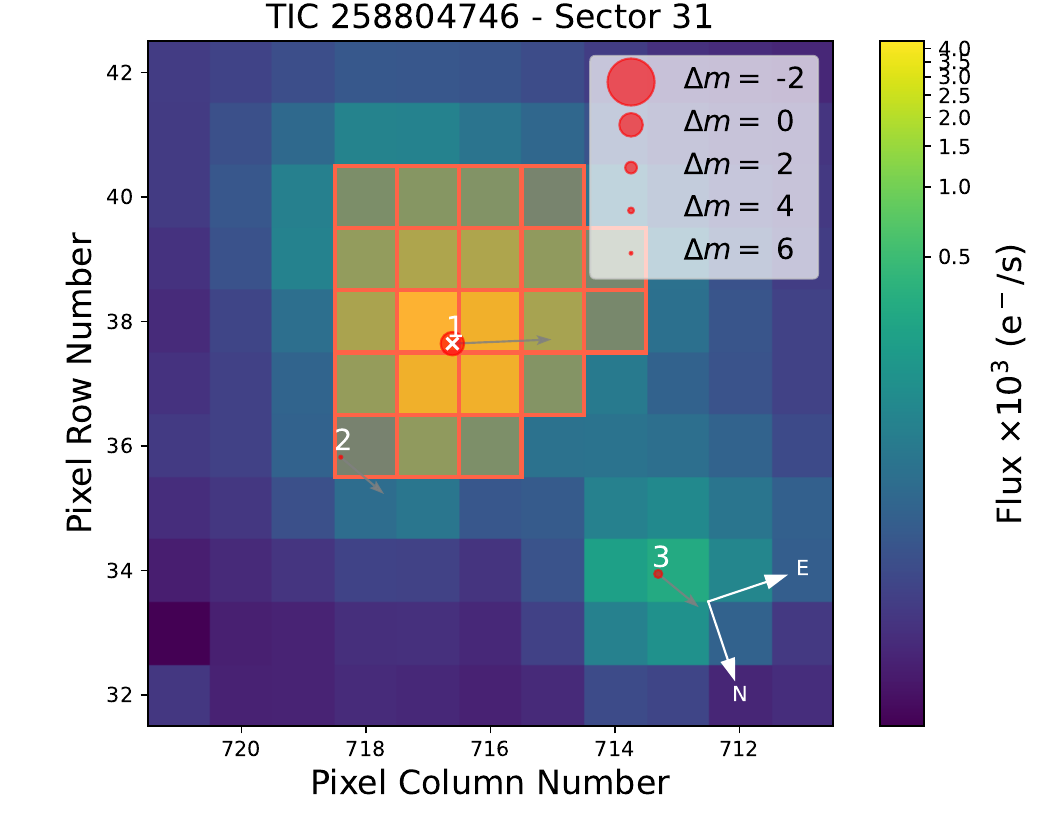}
\caption{TESS target pixel file image of TOI-2431 in Sector 31. The electron fluxes are shown with the color bar. The red-bordered pixels highlight the pixels used for the PDCSAP light curve. The TESS pixel scale is approximately $21"$. The size of the red circles indicates the TESS magnitudes of all nearby stars and TOI-2431 (white cross). Gray arrows show the proper motion directions of the stars in the field.}
\label{fig:TPF}
\end{figure}

TOI-2431 was observed by TESS in Sectors 31, 42, 43, 70, and 71 at a cadence of two minutes between 2020 and 2023. The TESS Science Processing Operations Center \citep[SPOC,][]{spoc} pipeline identified a small transit signal with a periodicity of $\sim$5 hours. We downloaded, and combined the available light curves using the \texttt{lightkurve} package \citep{2018ascl.soft12013L}. We worked with the Presearch Data Conditioning Simple Aperture Photometry (PDCSAP) SPOC light curves, which are corrected for pointing and focus-related instrumental signatures, discontinuities resulting from radiation events in the CCD detectors, outliers, and flux contamination. Figure~\ref{fig:TPF} shows the TESS pixels used for the PDC light curve generated with the \texttt{tpfplotter} code\footnote{\url{https://github.com/jlillo/tpfplotter}}. The TESS light curve, along with the best transit model, is shown in Section~\ref{sec:FCO}.

% ---------------------------------------------------
\subsection{Radial Velocities with NEID and HPF}
To constrain the mass of TOI-2431 b, we obtained high precision RVs using the NEID spectrograph \citep{schwab2016} on the WIYN 3.5m Telescope\footnote{The WIYN Observatory is a joint facility of the NSF National Optical-Infrared Astronomy Research Laboratory, Indiana University, the University of Wisconsin-Madison, Pennsylvania State University, and Princeton University.} at Kitt Peak National Observatory in Arizona. In total, we obtained 12 NEID exposures between December 17, 2024, and February 19, 2025. The exposure time for all observations was 5 minutes. The S/N varied per observation, with the median S/N of $27$ at $860 \, \rm{nm}$.

The NEID spectra were first extracted using the NEID
Data Reduction Pipeline \citep[DRP\footnote{\url{https://neid.ipac.caltech.edu/docs/NEID-DRP/}};][]{NEIDDRP}. These spectra were downloaded from the NEID archive\footnote{\url{https://neid.ipac.caltech.edu/search.php}}. We then extracted radial velocities from NEID using the \texttt{NEID-SERVAL} code, a spectral-matching code that builds on the SpEctrum Radial Velocity AnaLyzer \citep[SERVAL;][]{Zechmeister2018-do} code and which has been adapted for NEID data \citep{Stefansson2022-bw}. From this, we obtain a median RV precision of 2.9 $\mathrm{m\,s^{-1}}$.

In addition to the NEID data, to constrain the stellar parameters of the host star, we also obtained spectra of TOI-2431 with the Habitable-zone Planet Finder (HPF) Spectrograph \citep{Mahadevan_2012} on the 10m Hobby-Eberly Telescope \citep[HET,][]{1988ESOC...30..119R, Hill2021-os} at McDonald Observatory in Texas. We obtained three observations between December 29, 2024, and February 9, 2025, using $\sim$15 minute exposures. For these observations, the median S/N is $205$ at $1\,\mu m$ with the median RV precision of 7.3 $\mathrm{m\,s^{-1}}$.

The HPF 1D spectra were first reduced and extracted using the HPF pipeline, following the procedures outlined in \cite{10.1117/12.2312787}, \cite{2019ASPC..523..567K}, and \cite{2019Optic...6..233M}. We extracted the radial velocities using \texttt{HPF-SERVAL} following \cite{Stefansson2020_feb} and \cite{stefansson2023}. The RVs of TOI-2431 are listed in Table~\ref{table:rvobservations}.

% ---------------------------------------------------
\subsection{Speckle Imaging}
\begin{figure}
\centering
\includegraphics[width=\hsize]{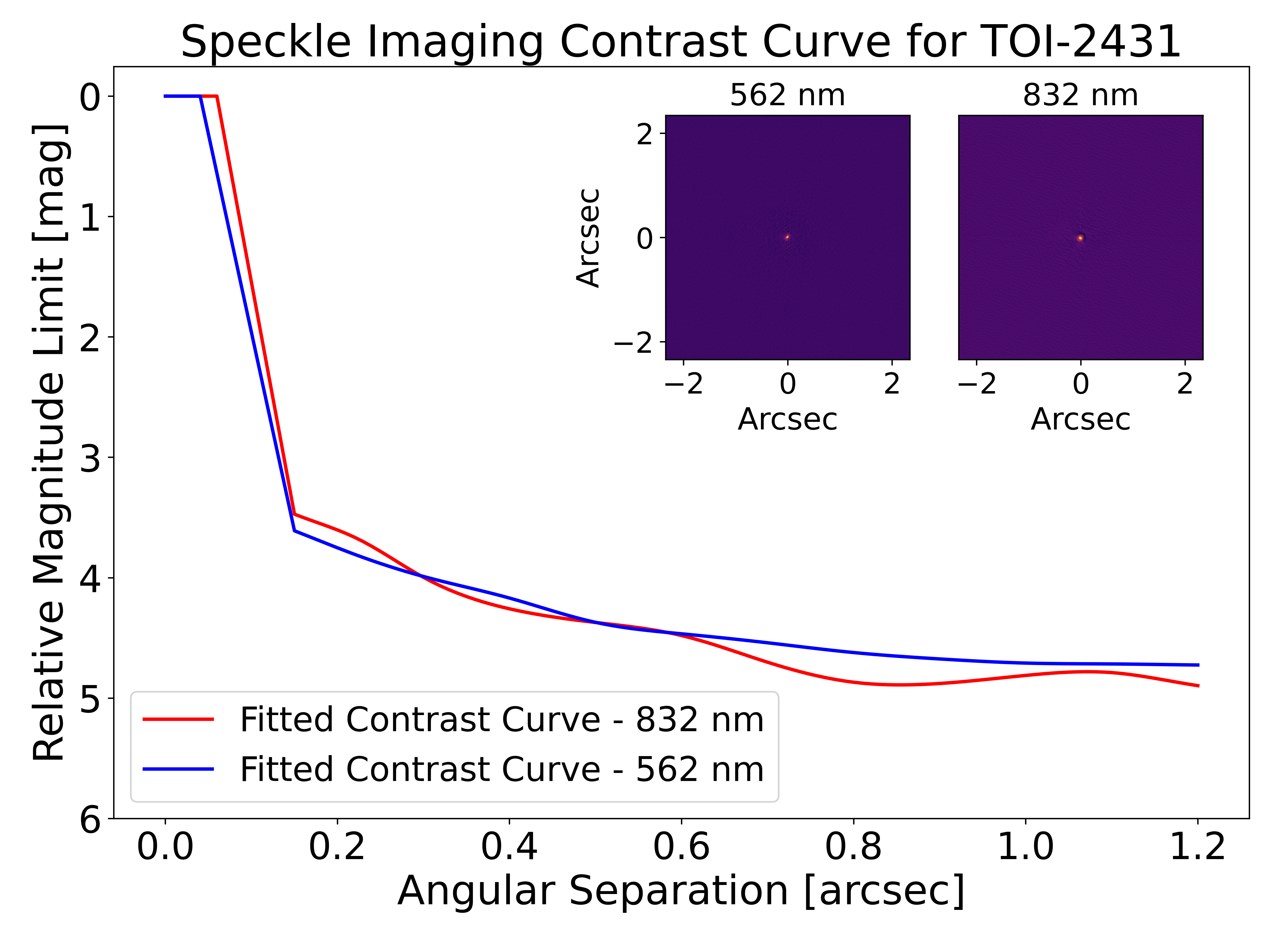}
\caption{Results from NESSI speckle imaging of TOI-2431. The plot shows the contrast curve as observed in the NESSI 562nm filter (blue curve), and the 832nm filter (red). The inset images highlight $256 \times 256$ insets of the reconstructed images. No secondary companions are seen. Note that in this figure, 1\arcsec corresponds to 36 AU at the distance of TOI-2431 b.}
\label{fig:Speckle_Imaging}
\end{figure}

To rule out the presence of nearby stars to TOI-2431, we observed TOI-2431 using the NESSI speckle imager \citep{Scott2018-ug} on the WIYN 3.5m Telescope at Kitt Peak in Arizona in two bands: the 562 nm band (width of 44 nm), and the 832 nm band \citep[width of 40 nm;][]{Scott2018-ug} available on NESSI. Figure~\ref{fig:Speckle_Imaging} shows the reconstructed images in both bands along with the corresponding $5\sigma$ contrast curve. The figure shows that no secondary sources were detected near (between 0.2\arcsec and 1.2\arcsec) the host star.

\begin{figure}
\centering
\includegraphics[width=\hsize]{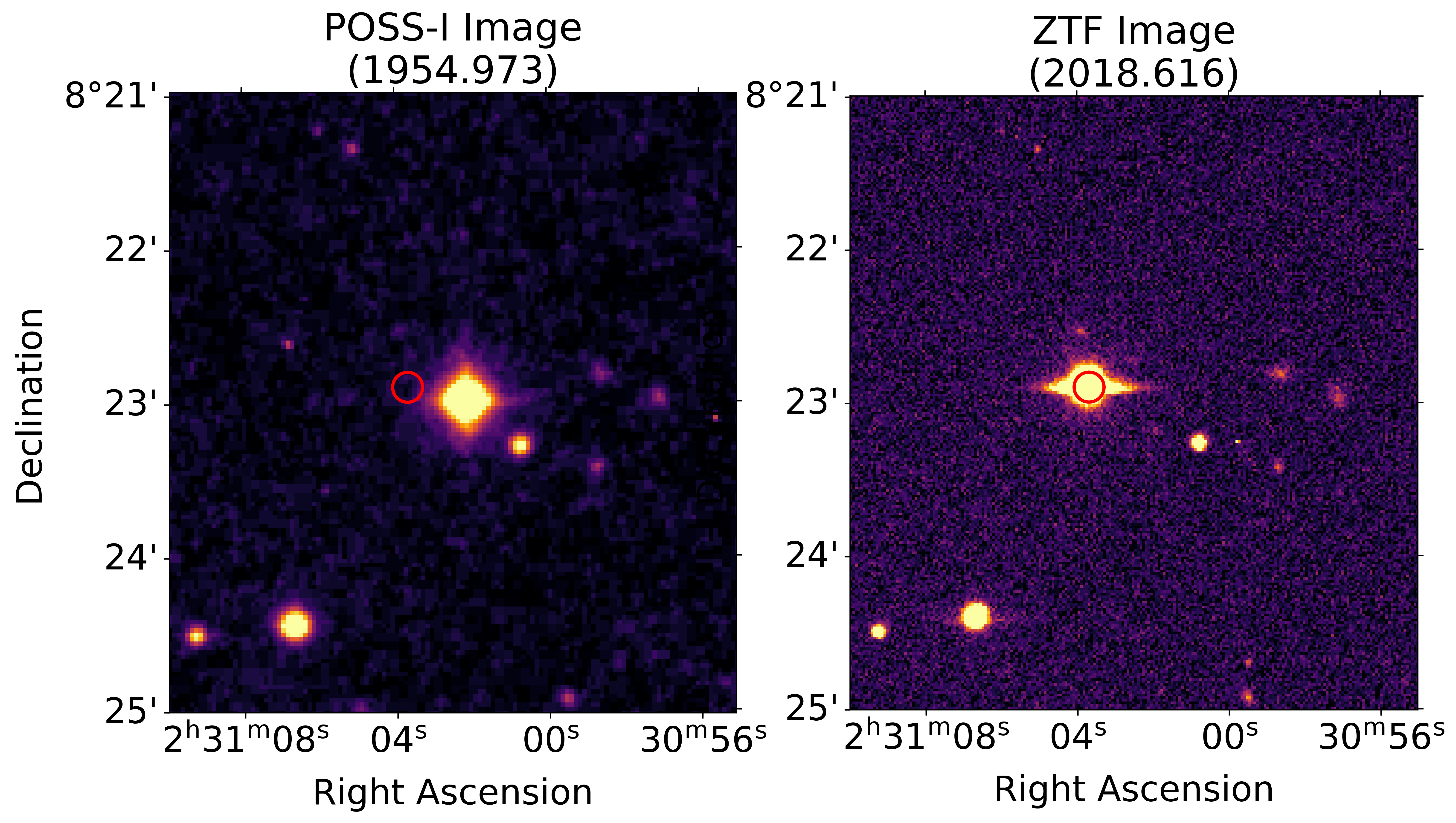}
\caption{Movement of TOI-2431 as a function of time. The left image shows a POSS-I image of TOI-2431 (middle star) taken in 1954, and the right image shows a ZTF image of TOI-2431 taken in 2018. The red circle highlights the position of TOI-2431 as measured by Gaia at the Gaia J2016 epoch. Due to its high proper motion, TOI-2431 has moved substantially between the two images, highlighting that there is no background star seen in the vicinity of TOI-2431 that could confound the planet interpretation of TOI-2431 b.}
\label{fig:Positional_shift}
\end{figure}

TOI-2431 is a high proper motion star, with a total proper motion of $384.5 \mathrm{\,mas \, yr^{-1}}$ \citep[$\mu_\alpha = 374.9 \mathrm{\,mas \, yr^{-1}}$, $\mu_\delta = -85.7 \mathrm{\,mas \, yr^{-1}}$;][]{Collaboration_Vallenari_Brown_Prusti_Bruijne_Arenou_Babusiaux_Biermann_Creevey_Ducourant_etal._2023}. Thus, to rule out any background star contamination as TOI-2431 moves across the sky, we also inspected the movement of TOI-2431 as a function of time, as seen in Fig.~\ref{fig:Positional_shift}. For this, we have used the First Palomar Sky Survey \citep[POSS-I;][]{POSS-I} image taken in 1954 and an image taken by the Zwicky Transient Facility \citep[ZTF;][]{masci2019} in 2018. To compare the images, we have also highlighted the position of TOI-2431 as measured by Gaia \citep{Collaboration_Vallenari_Brown_Prusti_Bruijne_Arenou_Babusiaux_Biermann_Creevey_Ducourant_etal._2023} at the Gaia J2016 epoch. The comparison showed that TOI-2431 has moved substantially from 1954 to 2018, and there are no background stars around it that could affect our interpretation.

% ------------------------------------------------------
% ------------------------------------------------------
% ------------------------------------------------------
\section{Stellar Parameters} \label{sec:stellarparams}
% ------------------------------------------------------
\subsection{Spectroscopic and Model Dependent Stellar Parameters}
We used the \texttt{HPF-SpecMatch} code \citep{Stefansson2020_feb} to constrain the stellar effective temperature ($T_{\mathrm{eff}}$), surface gravity ($\log{g}$), and metallicity ($\left[\mathrm{Fe/H}\right]$) of TOI-2431. \texttt{HPF-SpecMatch} compares an observed spectrum of a target star with a library of high S/N HPF spectra of well-characterized stars. For the analysis, we follow  \cite{Stefansson2020_feb}, and use the 5th HPF spectral order, which spans 8534–8645 $\AA$, as this order is the HPF order that is minimally impacted by telluric and sky-emission line contamination. Using \texttt{HPF-SpecMatch}, we obtain $T_{\mathrm{eff}} = 4080 \pm 77 \, \rm{K}$, $\log{g} = 4.68 \pm 0.05$, and $\left[\mathrm{Fe/H}\right] = -0.02 \pm 0.13 \, \rm{dex}$, where the uncertainties are obtained with the cross-validation approach outlined in \cite{Stefansson2020_feb} and \cite{jones2024}.

To obtain model-dependent estimates of the other stellar parameters of interest---including the mass, radius, and age of the host star---we made use of the \texttt{ARIADNE} code \citep{2022MNRAS.tmp..920V} to fit the Spectral Energy Distribution (SED) of the host star. To fit the SED, \texttt{ARIADNE} uses the Bayesian Model Averaging technique on a suite of different possible stellar models. For the fit, we used the \texttt{Phoenix v2} \citep{Husser2013}, \texttt{BT-Settl} \citep{Allard2012}, \cite{1993KurCD..13.....K}, and \cite{Castelli2004} stellar atmospheric models. We have also placed informative priors on $T_{\mathrm{eff}}$, $\left[\mathrm{Fe/H}\right]$, and $\log g$, from the \texttt{HPF-SpecMatch} analysis above. Given the close proximity of the star, we assumed that the optical interstellar extinction was negligible and set the corresponding extinction parameter $A_\mathrm{v}$ to zero. Within \texttt{ARIADNE} we adopted the default \texttt{ARIADNE} priors on the stellar radius (broad uniform prior ranging from $0.05 \, \rm{R}_\odot$ to $100 \, \rm{R}_\odot$), and the distance to the system \citep[informative gaussian derived by][]{BJ2021}. We list the best-fit parameters obtained from the fit in Table~\ref{table:stellarparams}. The stellar age and isochrone mass are calculated by \texttt{ARIADNE} using the best-fit parameters and the photometric data as input for a MIST isochrone interpolation \citep{MIST}. Lastly, \texttt{ARIADNE} cross-matches the best fit effective temperature to the spectral type table described by \cite{MamajekTable} to find the spectral type of the star. 

% --------------------------------------------------
\subsection{Stellar activity}\label{sec:activity}
The S/N of the NEID spectra are not sufficient to recover the Ca II H \& K S-value and $\log{R'_{HK}}$ activity indicators, so we can not use these to estimate expected RV jitter due to magnetic activity \citep{Wright2005,Isaacson2010,Luhn2020a}. However, we note that none of H$\alpha$ or the three Ca II NIR triplet lines is seen to be in emission, indicating that this star is inactive. Further, the TESS photometry does not show evidence of rotational modulation, which is broad agreement with the lack of detection of a clear rotational broadening in the HPF and NEID spectra ($v \sin i_\star < 2\,\mathrm{km\,s^{-1}}$). Together, these support that TOI-2431 is a slowly rotating star for which we do not expect strong rotational modulation in RVs, especially not on timescales close to those of the periodicity of the planet ($\sim$5 hours).

% ------------------------------------------------------
% ------------------------------------------------------
% ------------------------------------------------------
\begin{table*}
\centering
\caption{Parameters of TOI-2431. The photometric parameters shown were queried by \texttt{ARIADNE} and used for the SED fit.}
\label{table:stellarparams}
\begin{tabular}{llcc}
\hline\hline
Parameter & Description & Value & Source  \\
\hline
\multicolumn{4}{c}{\textbf{Main Identifiers}} \\
Gaia DR3 Source ID & -& 22707874346819712& Gaia \\
TIC &TESS Input Catalog Identifier & 258804746& TIC\\
TOI&TESS Object of Interest Identifier& 2431&\\
 HIP& Hipparchos Identifier&  HIP 11707&HIP\\
2MASS&2MASS ID & J02310327+0822550& 2MASS  \\
\hline
\multicolumn{4}{c}{\textbf{Equatorial Coordinates and Proper Motion}} \\
$\alpha_{J2016}$ (RA) & Right Ascension (RA), ICRS epoch 2016.0&  02:31:03.68& Gaia \\
$\delta_{J2016}$ (Dec) & Declination (Dec), ICRS epoch 2016.0&  08:22:53.85& Gaia\\
$\mu_{\alpha}$ (mas yr$^{-1}$) & Proper motion (RA, mas yr$^{-1}$)& $374.86 \pm 0.024$& Gaia\\
$\mu_{\delta}$ (mas yr$^{-1}$) & Proper motion (Dec, mas yr$^{-1}$)& $-85.68 \pm 0.018$& Gaia\\
 Spectral type& -&K7V &This work\\
\hline
\multicolumn{4}{c}{\textbf{Photometric Magnitudes}} \\
$U$ (mag)& APASS Johnson U magnitude& $13.5$ & APASS  \\
$B_{\text{TYCHO}}$ (mag) & TYCHO B magnitude & $12.14 \pm 0.14$& TYCHO  \\
$B$ (mag) & APASS Johnson B magnitude & $12.23 \pm 0.01$& APASS  \\
$BP$ (mag) & Gaia DR3 BP magnitude & $11.120 \pm 0.003$& Gaia \\
$V_{\text{TYCHO}}$ (mag) & TYCHO V magnitude & $11.03 \pm 0.08$& TYCHO  \\
$V$ (mag) & APASS Johnson V magnitude & $10.89 \pm 0.01$& APASS  \\
$G$ (mag) & Gaia DR3 G magnitude& $10.325 \pm 0.003$& Gaia\\
$RP$ (mag) & Gaia DR3 RP magnitude& $9.458 \pm 0.004$& Gaia \\
$T$ (mag) & TESS magnitude & $9.521 \pm 0.006$& TESS  \\
$J$ (mag) & 2MASS J magnitude & $8.36 \pm 0.02$& 2MASS  \\
$H$ (mag) & 2MASS H magnitude & $7.73 \pm 0.05$& 2MASS  \\
$K_S$ (mag) & 2MASS $K_S$ magnitude & $7.55 \pm 0.02$& 2MASS  \\
WISE1 (mag) & WISE1 magnitude & $7.49 \pm 0.03$& WISE  \\
WISE2 (mag) & WISE2 magnitude & $7.55 \pm 0.02$& WISE  \\
\hline
\multicolumn{4}{c}{\textbf{Stellar Parameters directly Derived by SED fitting}} \\
% $T_{\text{eff}}(K)$ & Effective temperature & $4109^{+28}_{-27}$& This work$^2$\\
$T_{\text{eff}}\,(\rm{K})$& Effective temperature &  $4109^{+28}_{-27}$& This work\\
% $\log{g}$ (cgs) & Stellar gravity & $4.69^{+0.05}_{-0.06}$& This work$^2$  \\
$\log{g}$ (cgs) & Stellar gravity & $4.68\pm0.05$ & This work\\
 $[{\rm Fe/H}]\,(\rm{dex})$ & Stellar metallicity & $-0.02\pm0.13$&This work\\
\hline
\multicolumn{4}{c}{\textbf{Derived Stellar Parameters}}\\

% $[{\rm Fe/H}]$ & Stellar metallicity & $-0.04^{+0.15}_{-0.12}$& This work$^2$  \\
$M_{\star, \text{iso}}\,(\rm{M_\odot})$& Stellar (Isochrone) Mass & $0.661^{+0.021}_{-0.024}$& This work\\
 $R_\star\,(\rm{R_\odot})$& Stellar radius & $0.651\pm 0.012$&This work\\
 $L_\star (L_\odot)$& Luminosity& $0.109\pm 0.005$&This work\\
$A_v$ (mag) & Visual extinction & $0.0$ (Fixed) & This work\\
$d$ (pc) & Distance & $36.01^{+0.06}_{-0.02}$& This work\\
 $\omega$ (mas)& Parallax& $27.76^{+0.02}_{-0.03} \, \mathrm{mas}$&This work\\
Age (Gyr) & Stellar age & $2.0^{+9.1}_{-1.7}$& This work\\
$\rho_\star \, (\rm{g \, cm^{-3}})$ & Stellar density& $3.23^{+0.18}_{-0.18} $&This work\\
\hline
\multicolumn{4}{c}{\textbf{Other Stellar Parameters}} \\
$v \sin i_\star$ (km s$^{-1}$) & Rotational velocity & $< 2$& This work  \\
% RV (km s$^{-1}$) & Radial velocity & X $\pm$ Y & This work  \\
RUWE & Gaia Renormalized Unit Weight Error & 1.070& Gaia \\
\hline
\end{tabular}
\tablefoot{References: Gaia (\citeyear{Gaia_DR1, Gaia_DR2, Collaboration_Vallenari_Brown_Prusti_Bruijne_Arenou_Babusiaux_Biermann_Creevey_Ducourant_etal._2023}), TESS \citep{2019AJ....158..138S,2018AJ....156..102S}, 2MASS \citep{2MASS}, WISE \citep{WISE}, APASS \citep{APASS}, TYCHO \citep{2000A&A...355L..27H}, HIP \citep{1997A&A...323L..49P}}
\end{table*}

% ------------------------------------------------
% ------------------------------------------------
% ------------------------------------------------
\section{Joint Transit and RV Analysis} \label{sec:analysis}

% ------------------------------------------------
\subsection{Joint Fits of Transit and RV Observations}
To confirm the planetary nature of TOI-2431 b and to better constrain its parameters, we performed joint fits of all of the available transit and RV data using the \texttt{juliet} code \citep{2019MNRAS.490.2262E}. \texttt{juliet} uses the \texttt{batman} code \citep{Kreidberg2015-oy} for the transit model, \texttt{radvel} \citep{Fulton2018-pq} for the RV model, \texttt{celerite} \citep{celerite} for the Gaussian Process (GP) modelling, and \texttt{dynesty} \citep{Speagle2019} for the dynamic nested sampling. We assumed that TOI-2431 b has a circular orbit ($e=0$), as we expect the orbit to circularize on a short timescale due to tidal interactions with its host star. Additionally, as the TESS Contamination Ratio is low at $0.0011$ \citep{spoc}, we assume no additional dilution of the transit due to nearby stars, and fix the dilution factor for all TESS photometry to unity in all of our fits. We have also tried TESS-Gaia Light Curve \citep[TGLC;][]{han2023tess} fit, which resulted in transit depths that are consistent within $1\sigma$ of the TESS light curve fit. To account for correlated noise in the TESS data, in all of the fits, we used a GP model with a Matern-3/2 kernel as implemented in the \texttt{celerite} code \citep{celerite}. Additionally, for the joint fits, we placed an informative prior on the stellar density of $\rho_\star = 3.230 \,\pm\,0.180 \, \rm g/cm^3$ (see Table~\ref{table:stellarparams}).

As a separate test, we performed similar fits with two additional codes: with the \texttt{ironman} code \citep{Espinoza-Retamal2023b,Espinoza-Retamal2024-jc}, and with the \texttt{ExoMUSE}\footnote{\url{https://github.com/Kaya-Han-Tas/ExoMUSE}} code, a code that we have developed that uses \texttt{radvel} and \texttt{batman} but uses Markov-Chain Monte-Carlo sampling using the \texttt{emcee} package \citep{emcee} instead of nested sampling. All fits resulted in consistent parameters within $\sim1\sigma$, and we elected to adopt the parameters we obtained using the \texttt{juliet} code.

We tested fitting the available NEID RV data assuming a single Keplerian, using informative priors on the orbital period of the planet and the transit midpoint from the TESS data. However, this resulted in a residual RV RMS scatter of 9.2 $\mathrm{m\,s^{-1}}$ that is substantially higher than that expected from the median NEID RV uncertainty of 2.9 $\mathrm{m\,s^{-1}}$. Since the star shows no clear signs of activity and USP planets are often found in multi-planet systems \citep{winn2018kepler}, this may suggest the presence of additional non-transiting planets in the system.

As highlighted in Section~\ref{sec:additionalplanets}, we looked for evidence of additional planets in both the TESS data using a box-least-squares analysis, and in the RVs using Generalized Lomb-Scargle (GLS) periodograms, but did not find statistically significant periodic signals in the currently available datasets (Section~\ref{sec:additionalplanets} for further details). As highlighted in Section~\ref{sec:additionalplanets}, we attribute the lack of periodic signals detected in the GLS analysis of the RVs due to the low total number of RVs (12 RVs), and urge additional RV observations to characterize any periodicities in the RV data.

To constrain the mass of TOI-2431 b, we used joint transit-and RV fits using two different approaches to account for the excess RV scatter: a) a joint transit and RV fit using the Floating Chunk Offset (FCO) method for the RVs, and b) a joint transit and RV fit employing both i) a Keplerian for the innermost planet, and ii) a quasi-periodic \textit{Gaussian Process (GP)} noise model to account for the excess RV scatter at longer periodicities than the USP. We discuss the results from the two fits below.

% ---------------------------------------------------
% -------------------------------------------------------
\subsubsection{Joint fit employing the Floating Chunk Offset (FCO) Method}\label{sec:FCO}
As USP planets tend to be found in multi-planetary systems, which could cause the RV data to be scattered in excess of the USP RV orbit, an often-used technique to account for the additional RV scatter is the \textit{Floating Chunk Offset (FCO)} method. The method was originally used by \cite{Hatzes2010-dh} to measure the mass of CoRoT-7b, an USP planet with a 0.85 day period, and since then, has been used to measure the masses for multiple short-period planets in the presence of other signals \citep{Hatzes_2014,Deeg2023-ch,Dai2024-qd}. 

The FCO method requires multiple observations to be taken over a single night spanning orbital phases that cause measurable velocity differences. It assumes a Keplerian RV orbit for the innermost planet and fits independent RV offsets between successive visits. Therefore, the method assumes that the only RV motion between successive visits (i.e., within a night) is dominated by the Keplerian model, and that the RV impact of other planetary signals are effectively removed by fitting the RV offsets. For TOI-2431 b, since the period is only 5.4 hours, we obtained 8 RV observations in three different visits, which had typical time-baseline of $\sim$1-2 hours within a visit, with a total time-baseline of $\sim$64 days.

We further note that although we do not expect significant RV variability on the rotational time scale for TOI-2431 given that it is an inactive star (Section~\ref{sec:activity}), the FCO method will also remove any activity-driven RV signals at the stellar rotation period.  Further, as TOI-2431 is an inactive star, we expect any sources of stellar variability within a night (e.g., oscillations, granulation) that would be retained via the FCO method are also expected to be small, and much less than the median RV uncertainty of the NEID data of 2.9 $\mathrm{m\,s^{-1}}$.

We implemented the FCO method in our joint fits by defining different nights of observations as different "instruments" that had different systematic velocities called NEID1, NEID2, and NEID3. We tried letting the jitter terms float, however, as the RV variation between the different nights is effectively removed by the different RV offsets, the jitter terms had posteriors consistent with zero. As both different runs resulted in the same planet parameter posteriors, we elected to adopt the simpler fit that fixed the jitter terms to zero.

As we only had three NEID visits to include for the FCO fit, we performed systematic injection and recovery tests to test the robustness of our FCO implementation, which we highlight in Appendix~\ref{sec:planetinjections}. In short, we took our observed NEID timestamps, and injected 1,500 synthetic RV signals of two planets: of a USP planet with the same ephemerides as TOI-2431 b, and another longer-period planet with a period in excess of one day. As highlighted in Appendix~\ref{sec:planetinjections}, the known RV amplitude is recovered at the rate as expected, we conclude that our FCO implementation and our mass constraint of TOI-2431 b is robust. 

\begin{figure*}
\centering  
\includegraphics[width=\hsize]{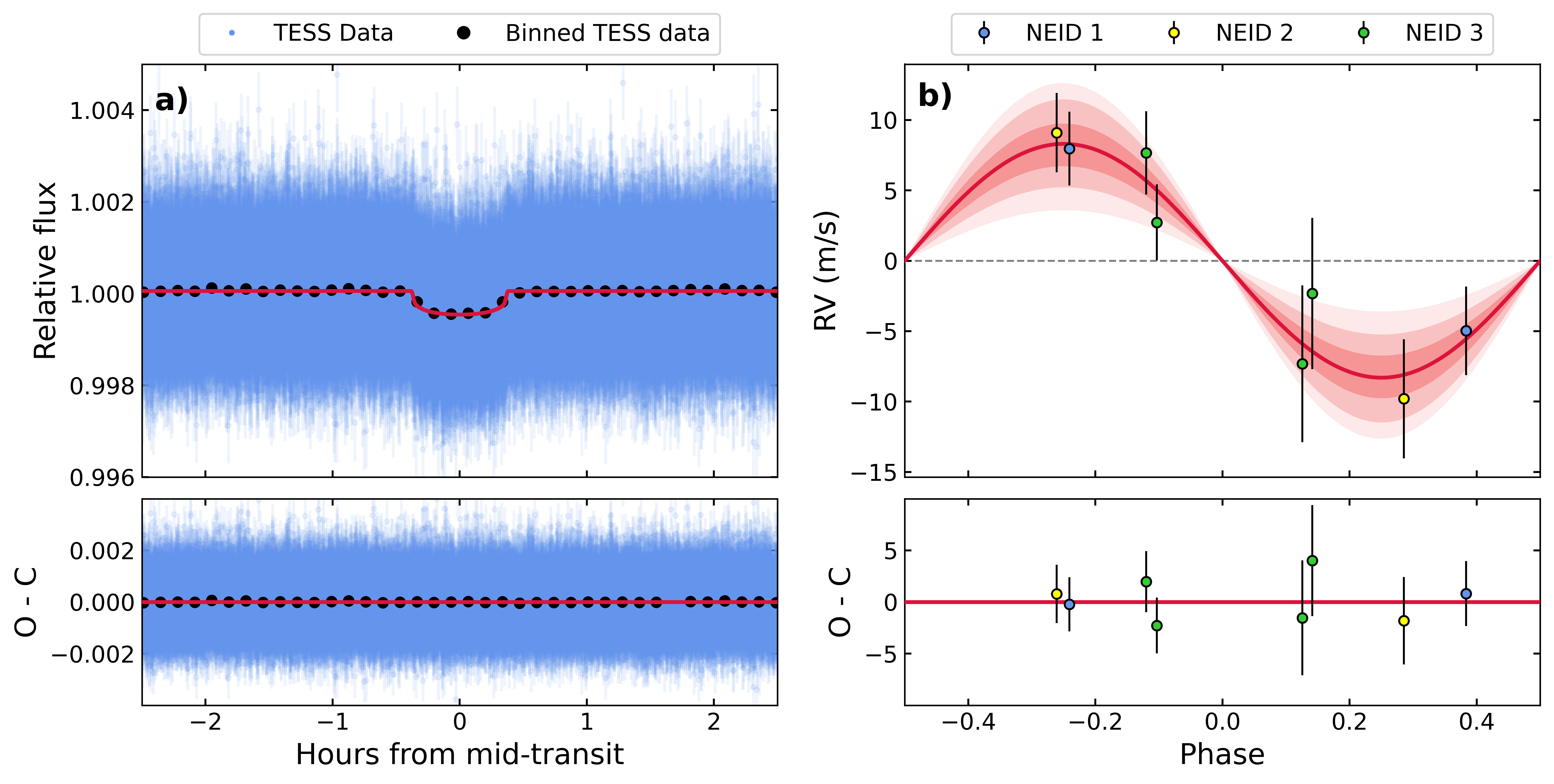}  
\caption{Best fit results from a joint transit and RV fit using the Floating Chunk Offset (FCO) method for the RV data. \textbf{a)} \textit{Top}: Detrended TESS photometric observations (2 min exposures) are shown in blue folded on the time of transit. Best-fit transit model is shown in red. The black points show the TESS data binned to a cadence of 2 min. \textit{Bottom:} Residuals from the best-fit model. \textbf{b)} \textit{Top:} Phase-folded RV data and best-fit model (red) of TOI-2431 as observed with NEID in three separate visits listed as NEID1 (blue), NEID2 (yellow), and NEID3 (green). The red shaded regions show the $1$, $2$, and $3\sigma$ credible intervals of the best-fitting model. \textit{Bottom:} Residuals from the best-fit model. We adopt the values from this fit as the values for TOI-2431 b.}
\label{fig:jointfit}  
\end{figure*}

The final joint fit is shown in Figure~\ref{fig:jointfit}, and the resulting best-fit parameters and associated priors are shown in Table~\ref{table:planetaryparams}.

% ---------------------------------------------------
\subsubsection{Joint fit employing a Quasi-periodic Gaussian Process}
\begin{figure*}
\centering 
\includegraphics[width=\hsize]{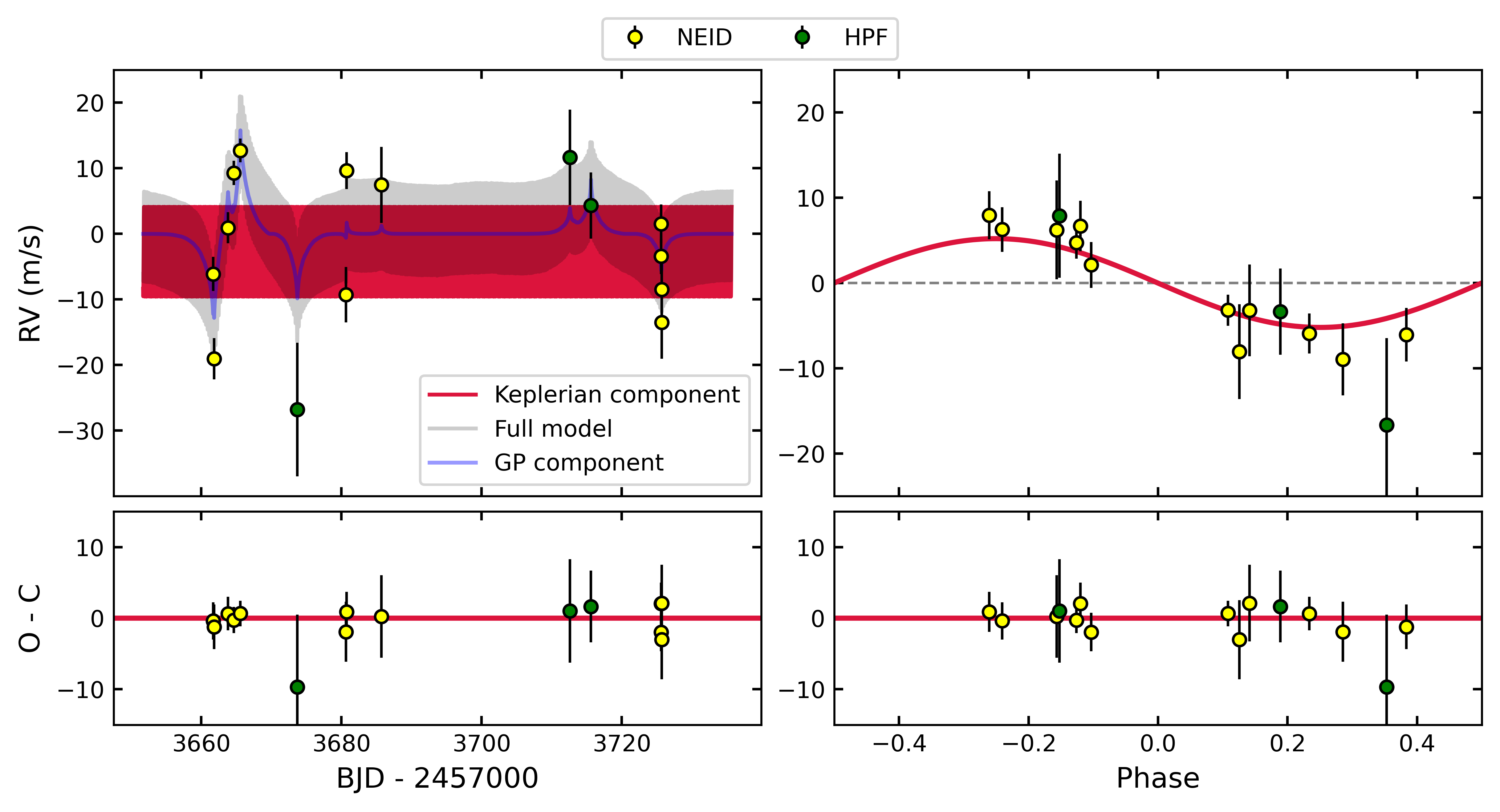}
\caption{Best fit results from a joint transit and RV fit using a quasi-periodic GP model to account for the excess RV scatter in the RVs. \textbf{a)} \textit{Top:} NEID (yellow) and HPF (green) RVs as a function of time. The best-fit Keplerian model is shown in red (appears as a solid band due to the short period orbit). The median of the GP component is shown in blue. The joint best-fit model is shown in grey. \textit{Bottom:} Residuals from the best fit model. \textbf{b)} \textit{Top:} Phase-folded RVs on the period of TOI-2431 b. Best fit Keplerian model after subtracting the GP model is shown in red. \textit{Bottom:} RV residuals as a function of phase.}
\label{fig:GPfit}  
\end{figure*}

In addition to the FCO fit discussed above, we performed a joint transit and RV fit where we used a Gaussian Process (GP) noise model with a quasi-periodic kernel to account for the excess RV scatter and the possible longer-term RV signals. To do so, we placed a uniform prior on the periodicity of the quasi-periodic GP from 1 to 100 days, to ensure it was substantially longer than the $\sim$5h period of the Keplerian USP component of the model. To provide the best constraint on the mass, for this fit, we used all of the available NEID RVs along with the HPF RVs. Here we note that NEID RVs are more precise than HPF since NEID has higher RV information content on bright K dwarfs than HPF. We let the instrument RV jitters for both instruments float. For consistency, the rest of the priors were kept the same as for the FCO joint fit priors. Table~\ref{table:planetaryparams} lists the priors and the resulting posterior parameters from the fit, and the resulting GP fit is shown in Figure~\ref{fig:GPfit}.

% ---------------------------------------------------

% ---------------------------------------------------
% ---------------------------------------------------
\section{Results}\label{sec:results}
The results from the FCO joint fit, and the GP joint fits are shown in Figures~\ref{fig:jointfit}, and~\ref{fig:GPfit}, respectively, and Table~\ref{table:planetaryparams} lists the resulting posteriors from both fits. 

\begin{table*}[!h]
\centering                         
\caption{Derived Parameters of TOI-2431 b from a joint analysis of the available transit and radial velocity data. $\mathcal{N}(\mu,\sigma)$ denotes a normal prior with mean ($\mu$) and standard deviation ($\sigma$),  $\mathcal{U}(a,b)$ denotes a uniform prior with a start value ($a$) and an end value ($b$), and $\mathcal{L}(\mu,\sigma)$ denotes a log-normal prior with mean ($\mu$) and standard deviation ($\sigma$). The posteriors from the Floating Chunk Offset (FCO) fit and the Gaussian Process (GP) fit are consistent.}
\label{table:planetaryparams}
\begin{tabular}{llccc}        
\hline\hline                 
 Parameter&Description &Prior Distribution& Posterior Distribution& Posterior Distribution\\
 & & & (FCO Fit, adopted) & (GP Fit)
\\
 \hline
\multicolumn{5}{l}{\textbf{MCMC Input Parameters:}} \\
$P$ $($days$)$& Orbital Period &$\mathcal{N}(0.224195,4\times10^{-7})$&\( 0.22419577_{-5\times10^{-8}}^{+5\times10^{-8}} \)&\( 0.22419578_{-5\times10^{-8}}^{+5\times10^{-8}} \)\\
\(T_0\) $($BJD$)$& Time of Transit (BJD - 2460000)&$\mathcal{N}(258.8689,4.6\times10^{-4})$&\( 258.86855_{-0.00015}^{+0.00015}\)&\( 258.86855_{-0.00015}^{+0.00016}\)\\      
\(e\)&Eccentricity &\(0\) (Adopted)& \(0\) (Adopted) &\(0\) (Adopted) \\
\({\omega}\) $($deg$)$&Argument of Periapsis &\(90\) (Adopted)& \(90\) (Adopted) &\(90\) (Adopted) \\
\({b}\)&Impact Parameter &$\mathcal{U}(0,1)$& \( 0.574_{-0.031}^{+0.033}\)&\( 0.572_{-0.035}^{+0.033}\)\\
\(K\) $($m s$^{-1}$$)$&RV Semi-Amplitude  &$\mathcal{U}(0,1000)$&\( 8.3_{-1.6}^{+1.5} \)&\( 6.8_{-2.0}^{+2.1} \)\\
\(q_{1}\)&TESS Limb Darkening Coefficient&$\mathcal{U}(0,1)$&\( 0.55_{-0.18}^{+0.22} \) &\( 0.59_{-0.19}^{+0.23} \)\\
\(q_{2}\)&TESS Limb Darkening Coefficient&$\mathcal{U}(0,1)$&\( 0.24_{-0.16}^{+0.28} \) &\( 0.23_{-0.16}^{+0.26} \)\\
\(m_{\text{dilution, TESS}}\)& Dilution Factor of TESS& \(1\) (Adopted)&\(1\) (Adopted) &\(1\) (Adopted) \\
\(m_{\text{flux, TESS}}\)& Offset Relative Flux& $\mathcal{N}(0.0,0.1)$&\( -0.000057_{-0.000032}^{+0.000031} \)&\( -0.00006_{-0.00003}^{+0.00003} \)\\
$\rm GP_{\sigma, \, \text{TESS}}$ (ppm)& Amplitude of the GP&$\mathcal{L}(10^{-6},10^{6})$&\( 0.00027_{-0.000017}^{+0.000020} \)&\( 0.00027_{-0.000017}^{+0.000019} \)\\
$\rm GP_{\rho, \, \text{TESS}}$ (ppm)& Length-Scale of the Matern Kernel&$\mathcal{L}(10^{-6},10^{6})$&\( 0.68_{-0.078}^{+0.087} \)&\( 0.68_{-0.078}^{+0.084} \)\\
\({\sigma_{\text{TESS}}}\) (ppm)& Light Curve Jitter& $\mathcal{L}(10^{-6},10^{6})$&\( 177_{-10}^{+9.8} \)&\( 176_{-10}^{+9.8} \)\\
$\rm GP_{\text{B, RV}}$ $($m s$^{-1}$$)$& Amplitude of the GP (RV)& $\mathcal{L}(10^{-6},10^{6})$& $-$&\( 137_{-65}^{+179} \)\\
$\rm GP_{\text{C, RV}}$& Constant Scaling Term of the GP& $\mathcal{L}(10^{-6},10^{6})$& $-$&\( 0.40_{-0.40}^{+7620} \)\\
$\rm GP_{\text{L, RV}}$& Char. Time-Scale of the GP& $\mathcal{L}(10^{-3},10^{3})$& $-$&\( 1.6_{-1.3}^{+6.8} \)\\
$\rm GP_{\text{$P_{rot}$, RV}}$ (days)& Period of the Quasi-Periodic GP& $\mathcal{U}(1,100)$& $-$&\( 47_{-31}^{+35} \)\\
${R_p}/{R_*}$ & Radius Ratio &$\mathcal{U}(0,1)$&\( 0.02133_{-0.00031}^{+0.00033} \)&\( 0.02128_{-0.00032}^{+0.00036} \)\\
${\rho}_{*}$ $(\text{g}\,\text{cm}^{-3})$& Density of Star &$\mathcal{N}(3.23,0.18)$&\( 3.24_{-0.18}^{+0.18} \)&\( 3.24_{-0.18}^{+0.18} \)\\
\(v_{\gamma,\,\text{obs. log }}\) $($m s$^{-1}$$)$& Systematic Velocity (NEID, GP)& $\mathcal{U}(-50,50)$& $-$&\( -2.7_{-5.5}^{+5.4} \)\\
\(v_{\gamma,\,\text{HPF}}\) $($m s$^{-1}$$)$& Systematic Velocity (HPF, GP)& $\mathcal{U}(-50,50)$& $-$&\( 1.3_{-9.5}^{+8.9} \)\\
\(v_{\gamma,\,1}\) $($m s$^{-1}$$)$& Systematic Velocity (NEID 1)&$\mathcal{U}(-50,50)$&\( -17_{-2.1}^{+2.0} \)&$-$\\
\(v_{\gamma,\,2}\) $($m s$^{-1}$$)$& Systematic Velocity (NEID 2)& $\mathcal{U}(-50,50)$&\( -2.3_{-2.4}^{+2.4} \)&$-$\\
\(v_{\gamma,\,3}\) $($m s$^{-1}$$)$& Systematic Velocity (NEID 3)& $\mathcal{U}(-50,50)$&\( -9.0_{-1.8}^{+1.9} \)&$-$\\
\({\sigma_{\text{RV}, \,\text{NEID}}}\) $($m s$^{-1}$$)$& RV Jitter (NEID, GP)& $\mathcal{L}(10^{-6},10^{6})$& $-$&\( 0.0047_{-0.0046}^{+0.63} \)\\
\({\sigma_{\text{RV}, \,\text{HPF}}}\) $($m s$^{-1}$$)$& RV Jitter (HPF, GP)& $\mathcal{L}(10^{-6},10^{6})$& $-$&\( 0.0069_{-0.0069}^{+2.2} \)\\
\({\sigma_{\text{RV}, \,1}}\) $($m s$^{-1}$$)$& RV Jitter (NEID 1)& \(0\) (Adopted)&\(0\) (Adopted) &\(0\) (Adopted) \\
\({\sigma_{\text{RV},\, 2}}\) $($m s$^{-1}$$)$&  RV Jitter (NEID 2)& \(0\) (Adopted)&\(0\) (Adopted) &\(0\) (Adopted) \\
\({\sigma_{\text{RV},\,3}}\) $($m s$^{-1}$$)$&  RV Jitter (NEID 3)&\(0\) (Adopted)&\(0\) (Adopted) &\(0\) (Adopted) \\       
\multicolumn{5}{l}{\textbf{Derived Parameters:}} \\
\(M_{p}\) $(\rm{M_\oplus})$&Planet Mass &$-$&\( 6.2_{-1.2}^{+1.2}\)&\( 5.1_{-1.5}^{+1.6}\)\\
\(R_{p}\) $(\rm{R_\oplus})$& Planet Radius &$-$&\( 1.536_{-0.033}^{+0.033}\)&\( 1.533_{-0.033}^{+0.034}\)\\
\(\rho_{p}\) $(\text{g}\,{\text{cm}^{-3}})$& Planet Density &$-$&\( 9.4_{-1.8}^{+1.9}\)&\( 7.8_{-2.4}^{+2.5}\)\\
\(i\) (deg)&Inclination &$-$& \( 74_{-0.96}^{+0.99}\)&\( 74_{-1.0}^{+1.1}\)\\
\(a\) $(\text{AU})$& Semi-Major Axis & $-$&\( 0.0063_{-0.0001}^{+0.0001}\) &\( 0.0063_{-0.0001}^{+0.0001}\) \\
ESM& Emission Spectroscopy Metric &$-$&\( 27\)&\( 27\)\\     
\({T_{\text{eq, a = 0}}}\) $($K$)$& Equilibrium Temp. ($A_{\rm{B}} = 0$)& $-$& \( 2063_{-30}^{+30} \)&\( 2063_{-30}^{+30} \)\\
\({T_{\text{eq, a = 0.3}}}\) $($K$)$& Equilibrium Temp. ($A_{\rm{B}} = 0.3$)& $-$& \( 1887_{-28}^{+28} \)&\( 1887_{-28}^{+28} \)\\
\hline
\end{tabular}
\end{table*}

The FCO joint fit yields a RV semi-amplitude of $8.3_{-1.6}^{+1.5}\,\mathrm{m\,s^{-1}}$, and a mass of $6.2_{-1.2}^{+1.2}\,\rm{M_\oplus}$. In comparison, the GP joint fit yields an RV semi-amplitude of $6.8_{-2.0}^{+2.1} \, \mathrm{m\,s^{-1}}$, and a mass of $5.1_{-1.5}^{+1.6} \, \rm{M_\oplus}$. Both the RV semi-amplitude and mass values are consistent within their $1\sigma$ uncertainties, as are the other posterior values, as shown in Table~\ref{table:planetaryparams}. As the FCO fit is a simpler fit, with fewer free parameters (one parameter for the Keplerian and three RV offsets, compared to one Keplerian parameter, two RV offsets, two RV jitter terms, and four parameters for the GP), we elect to adopt the values for the FCO fit. This constrains the mass of the planet at $\sim5.2\sigma$ confirming TOI-2431 b as a USP planet.

From Table~\ref{table:planetaryparams}, we see that the FCO joint fit results in a planet radius of $1.536_{-0.033}^{+0.033} \, \rm{R_\oplus}$. This suggests a density of $9.4_{-1.8}^{+1.9}\,\mathrm{g \, cm^{-3}}$, which is slightly denser, but consistent with an Earth-like composition.

% -----------------------------------------------------------
% -----------------------------------------------------------
% -----------------------------------------------------------
\section{Discussion} \label{sec:discussion}

% -----------------------------------------------------------
\subsection{Search for Additional Planets}\label{sec:additionalplanets}
\begin{figure}[h]
\centering
\includegraphics[width=\linewidth]{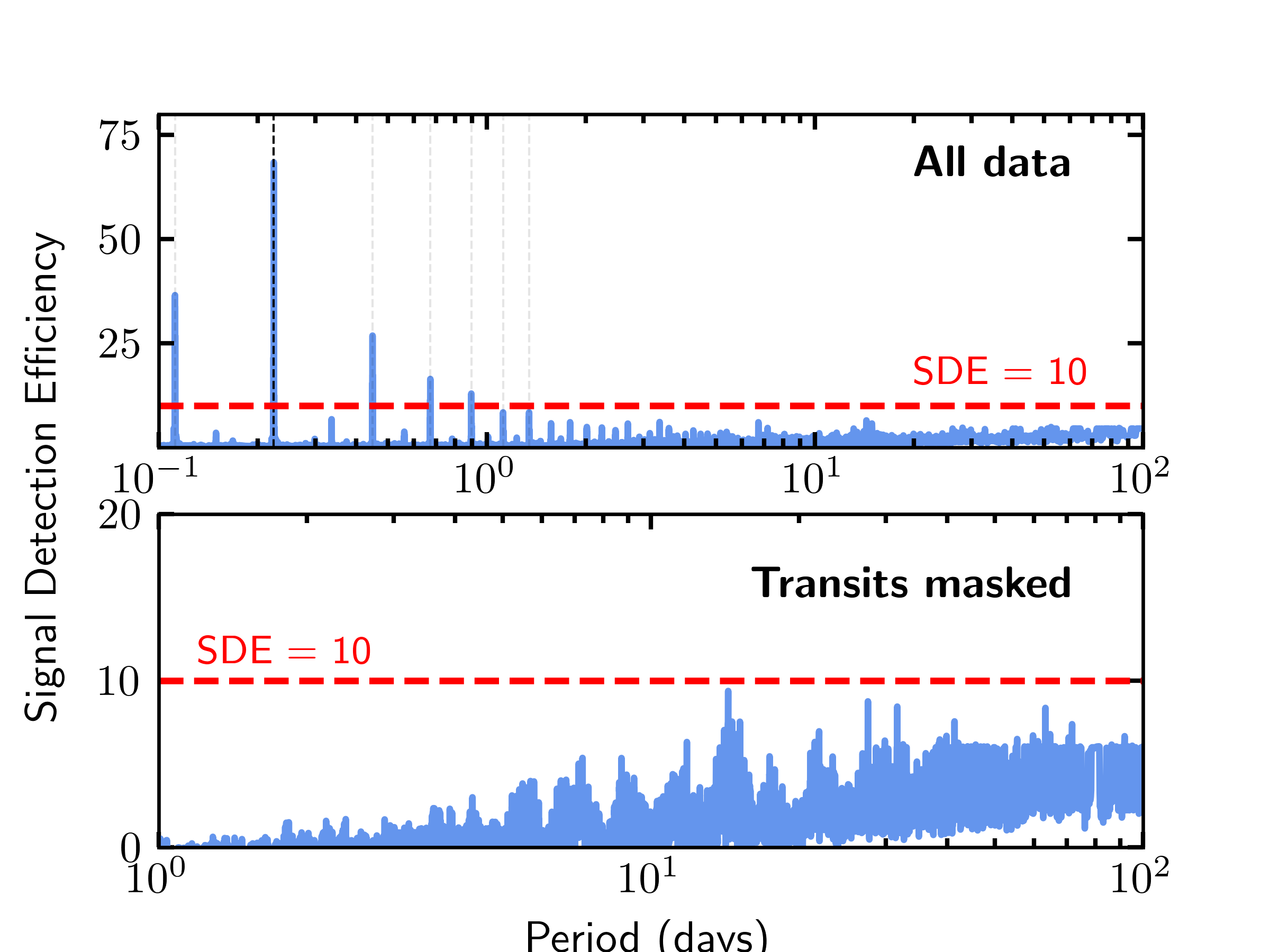}
\caption{Box-least square (BLS) periodogram of the TESS data. \textit{Top:} BLS periodogram of all of the TESS sectors. The highest peak reveals the period of TOI-2431 b of 0.22 days, which is highlighted with the black dashed vertical line. Aliases of this period are also shown as faint dashed lines. The horizontal red dashed line shows a Signal Detection Efficiency of 10, the threshold we adopt for a significant signal. \textit{Bottom:} BLS periodogram of the TESS data after masking out the transits of TOI-2431 b. No significant peaks are detected.}
    \label{fig:BLS}
\end{figure}

To search for evidence of additional transiting planets in the TOI-2431 system, we used the Box-Least Square (BLS) algorithm as implemented in the \texttt{lightkurve} package \citep{2018ascl.soft12013L}. The resulting BLS periodogram is shown in Figure~\ref{fig:BLS}. We do not observe any clear signal in the BLS other than the signals corresponding to TOI-2431 b and its aliases. Further, when removing that signal, no additional signals with a Signal Detection Efficiency (SDE) larger than 10 are detected. We see a potential signal with a period of $\sim14$ days and an SDE of $\sim9$, but we attribute this to the combined effect of the gaps in the TESS data (the signal is close to half of the TESS baseline per sector) and the masking of TOI-2431 b transits, which removed a significant number of data points. Therefore, we conclude there are no clear signs of additional transiting planets in the system. The lack of detection of other transiting planets is not unexpected, as the planets in USP systems tend to have a high mutual inclinations \citep{Dai_2018}, which together with the high period ratios between neighboring planets \citep{Dai_2015} can easily explain the non-detection of additional transiting planets in the system.

\begin{figure}
\centering
\includegraphics[width=0.9\linewidth]{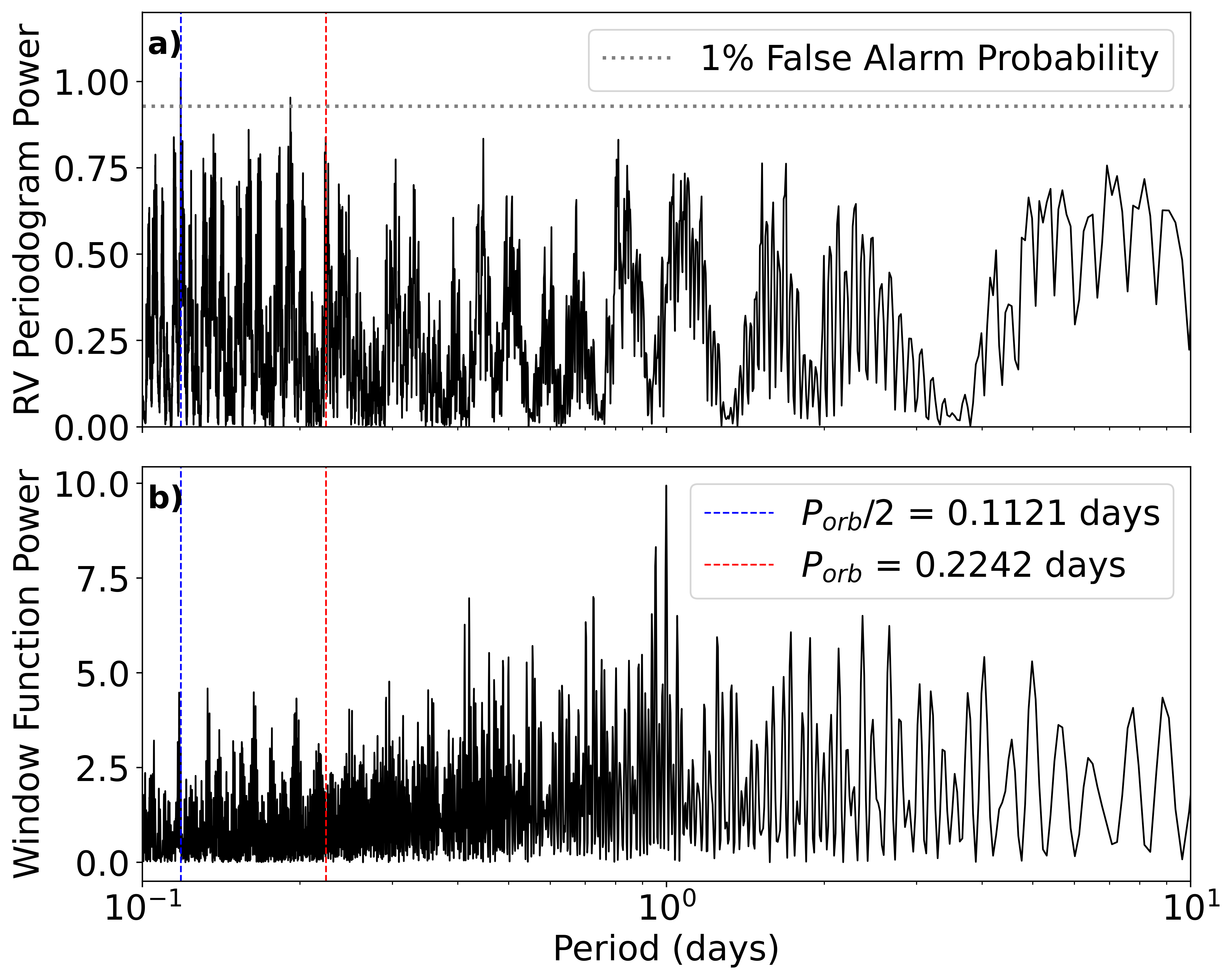}
\caption{Generalized Lomb-Scargle (GLS) Periodogram and Window Function of the NEID RVs of TOI-2431 \textbf{a)} GLS Periodogram. The grey dotted line represents the False Alarm Probability (FAP) of 1\%. The period of TOI-2431 b is highlighted with red vertical dashed line and the half-period is highlighted with blue vertical dashed line. Beyond the peak seen at $P_{\mathrm{orb}}/2$, we see no additional signals with FAP$<1\%$. \textbf{b)} Window function of the RVs in a).}
\label{fig:LSandWF}
\end{figure}

To look for evidence of non-transiting planets in the RVs, we show a Generalized Lomb-Scargle (GLS) Periodogram of the NEID RVs along with the accompanying window function in Figure~\ref{fig:LSandWF}. We highlight the orbital period of TOI-2431 b's with the red vertical dashed line, and we highlight half of the orbital period with the blue vertical dashed line. The latter has a False Alarm Probability (FAP) $<1\%$ (grey horizontal line), which we attribute being likely due to TOI-2431 b. The window function shows that these peaks do not originate from an artifact due to sampling. The grey dotted line shows that we do not detect any additional significant periodicities with FAP$<1\%$. We attribute this being due to the low number of RVs, as periodogram analyses like these benefit from higher number of datapoints, especially in the absence of extremely clear sinusoidal signals. Given the high RV scatter compared to expectations from a single Keplerian, we urge additional RV observations of the system to look for evidence of significant additional periodicities to reveal additional planets in the TOI-2431 system.

In addition to RVs, massive longer-period exoplanets might be detected using astrometry. The star is included in the Hipparcos-Gaia Catalog of Accelerations \citep{Brandt2021}. However, we do not see any significant proper motion differences, ruling out massive companions at large orbital distances. Nonetheless, future Gaia DR4 astrometry can put more constraints on possible outer companions at intermediate orbital distances of $\sim3-5$ AU \citep{Perryman2014}. The star is relatively nearby ($d\sim36$ pc) and bright ($G\sim10.3$), which increases the likelihood of detecting massive outer companions to TOI-2431 b with Gaia \citep[and possibly measuring the mutual inclination between the planetary orbits;][]{Espinoza-Retamal2023}.

\begin{figure*}
\centering
\includegraphics[width=0.85\hsize]{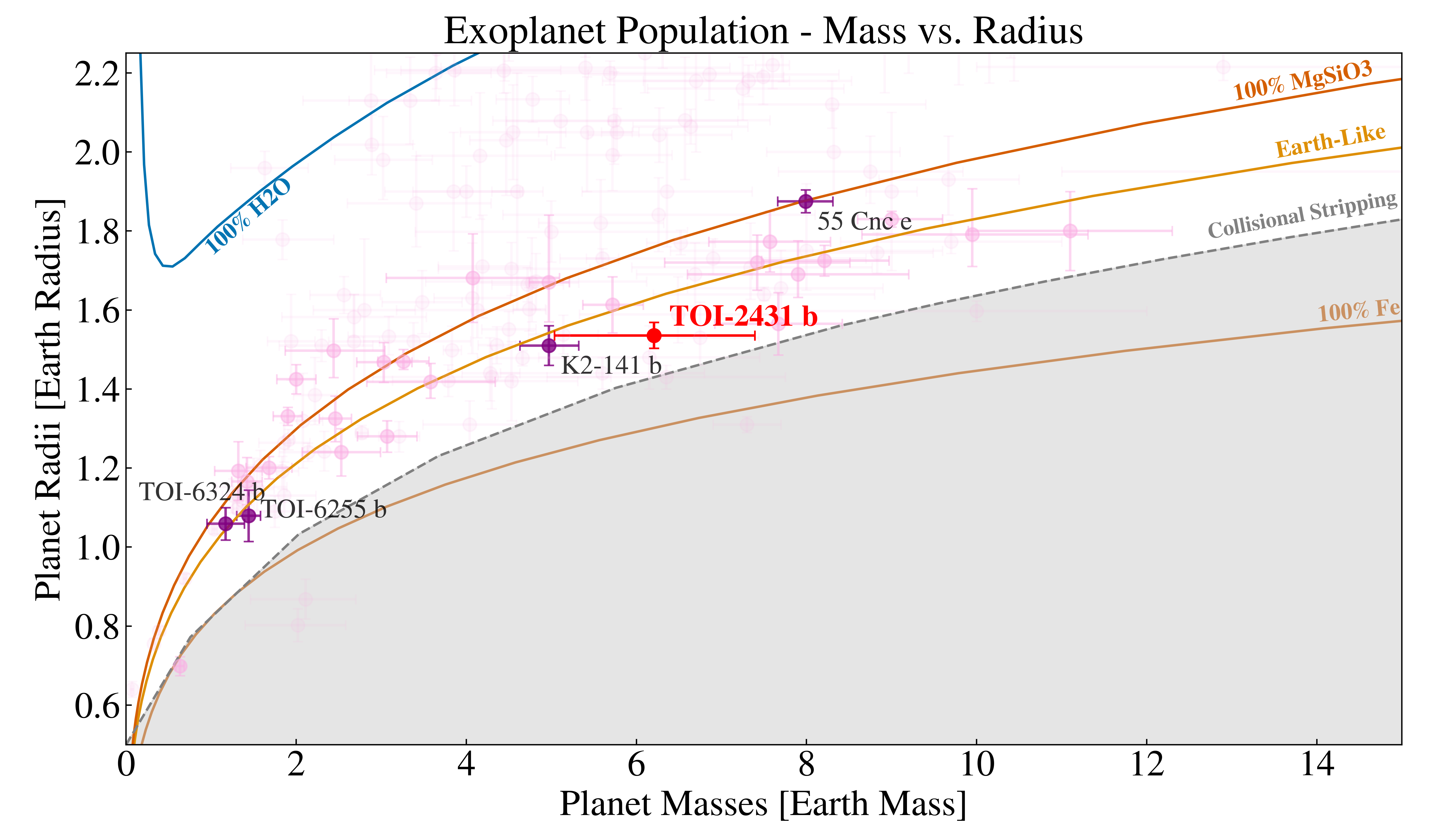}
\caption{Planet radius as a function of planet mass along with lines of constant density from \citep{doi:10.1073/pnas.1812905116}. The gray shaded region indicates planets with iron content exceeding the maximum value predicted from models of collisional stripping \citep{2010ApJ...712L..73M}. Confirmed USP planets are shown with bold foreground points, and other confirmed exoplanets from the NASA Exoplanet Archive are shown with the faint purple points. TOI-2431 b is shown in red. We only show systems with better than $3\sigma$ mass and radius constraints.}
\label{fig:densityplot}
\end{figure*}

% -----------------------------------------------------
\subsection{Planet Composition}
Figure~\ref{fig:densityplot} compares the mass and radius of TOI-2431 b compared to other USP planets (bold points) and other exoplanets (faint points) with well-characterized masses and radii. To examine possible compositions that TOI-2431 b is compatible with, we additionally plot planet composition models from \cite{doi:10.1073/pnas.1812905116} as lines of constant density. We see that TOI-2431 b is slightly denser, but compatible with an Earth-like planet composition, similar to other well-characterized USP planets. Due to its close distance to the host star and the high incident irradiation, this results in an equilibrium temperature of $2063\pm 30 \, \mathrm{K}$ (assuming Bond Albedo as $A_{\mathrm{B}}=0$). This temperature exceeds the melting points of many common silicate minerals found in planets, suggesting that the planet's surface is likely to be partially or fully molten.

% -----------------------------------------------------
\subsection{Star-Planet Tidal Interactions} \label{ssec:tidal}
The short orbital period of TOI-2431 b suggests that it is subject to strong tidal interactions with its host star, potentially leading to tidal deformation and orbital decay. We discuss these tidal effects further below, broadly following similar discussions of the USP planets TOI-6255 b \citep{Dai2024-qd}, and TOI-6324 b \citep{RenaLee_2025ApJ}.

\begin{figure}
\centering
\includegraphics[width=0.9\hsize]{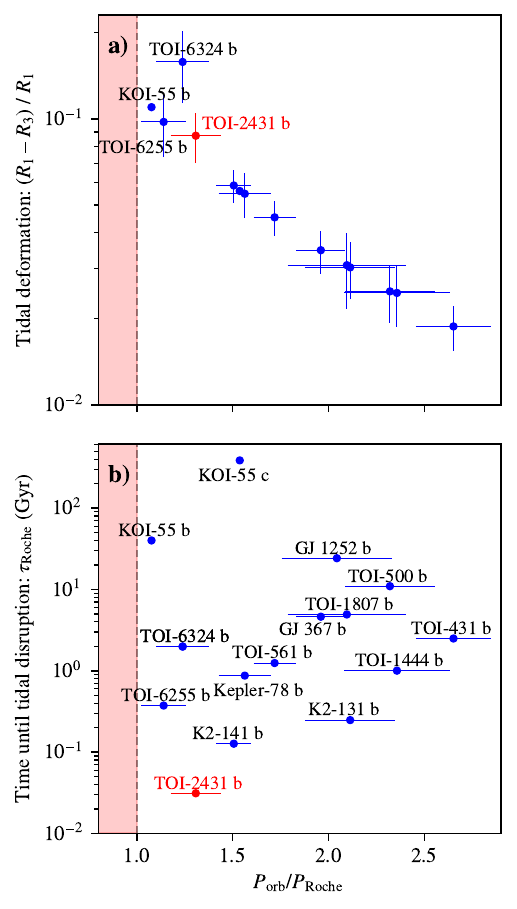}
\caption{\textbf{a)} Expected tidal deformation of USPs parametrized as a $(R_1 - R_3)/R_1$ as a function of $P_{\mathrm{orb}}/P_{\mathrm{Roche}}$. Here the planets are approximated as compressed ellipsoids with primary axes radii of $R_1$, $R_2$, and $R_3$, where $R_1$ is the radius of the planet pointing towards the star, and $R_3$ is the radius along the orbit normal. We see that TOI-2431 b is among some of the most tidally distorted USP planets. \textbf{b)} Roche limit timescales of USPs as a function of $P_{\mathrm{orb}}/P_{\mathrm{Roche}}$. The latter indicates the proximity to the Roche limit, and the former indicates the timescale to reaching the Roche limit. We see that TOI-2431 b has the shortest timescale until tidal disruption of $\sim$31 Myr.}
\label{fig:tidal}
\end{figure}

% ---------------------------------------------------
\subsubsection{Proximity to the Roche Limit and Tidal Deformation}
The minimum period at which a planet of mean density $\rho_{\rm p}$ can orbit before being disintegrated by tidal forces is set by the Roche limit. Approximating a planet as an incompressible fluid, this Roche period is given by \citep{Rappaport2013-ew}:
\begin{equation}
P_{\rm Roche} \approx 12.6 \,{\rm h} \, \left( \frac{\rho_{\rm p}}{1\,{\rm g\,cm^{-3}}} \right)^{-1/2}.
\label{eq:p_roche}
\end{equation}
Using TOI-2431 b's bulk density of $\rho_{\rm p} = 9.4_{-1.8}^{+1.9}$  g\,cm$^{-3}$, we estimate $P_{\rm Roche} = 4.11^{+0.48}_{-0.36}$ h. Therefore, $P_{\rm orb}/P_{\rm Roche} = 1.31^{+0.12}_{-0.14}$, suggesting that TOI-2431 b has an orbital period only $\sim$30\% larger than the Roche limit and that tidal forces are likely to produce measurable effects on the planet.

One possible effect of the proximity of TOI-2431 b to its host star is to tidal deformation, where the shape of the planet is elongated towards the host star. To quantify this effect, we can describe the planet as an elongated ellipsoid parametrized as having three different radii, $R_1$, $R_2$, $R_3$, where $R_1$ is the radius of the planet along the axis that points towards the host star, $R_2$ is along the direction of orbit, and $R_3$ is along the orbit normal. Following \cite{Dai2024-qd}, the radii are given by:
\begin{equation}
R_1=R_{\rm vol}(1+\frac{7}{6}\delta R_{p}),
\end{equation}
\begin{equation}
R_2=R_{\rm vol}(1-\frac{1}{3}\delta R_{p}),
\end{equation}
\begin{equation}
R_3=R_{\rm vol}(1-\frac{5}{6}\delta R_{p}),
\end{equation}
\noindent where $\delta R_{p}$ is the tidal distortion of the planet, and $R_{\rm vol}$ is the volumetric radius of the planet defined as $R_{\rm vol} \equiv (R_1R_2R_3)^{1/3}$. Given its short-period orbit, TOI-2431 b is expected to be tidally locked \citep{winn2018kepler}, with a rotational period equal to its orbital period, resulting in $R_1$ being hidden from the observer during transit. The radius that is observed during transit therefore is $R_{\rm tran} \equiv (R_2R_3)^{1/2}$, giving rise to the relation,
\begin{equation}
R_{\rm vol} = R_{\rm tran}(1+\frac{7}{12}\delta R_{p}).
\end{equation}

Following \cite{Dai2024-qd}, the tidal distortion $\delta R_{p}$ can be calculated using
\begin{equation}
\delta R_p = h_2 \zeta,
\end{equation}
\begin{equation}
\zeta = \frac{M_{\star}}{M_p} \left(\frac{R_p}{a}\right)^3,
\end{equation}
where $R_p$ is the planetary radius, $h_2$ is the Love number, $a$ is the semi-major axis, and $M_p$ and $M_{\star}$ are the masses of the planet and host star, respectively. Following \cite{Dai2024-qd}, we adopt a fiducial $h_2=1$, from which we find that TOI-2431 b's volumetric radius is ~2.4\% larger than the observed transit radius. Further, we find that the difference between the longest axis $R_1$ and the shortest axis $R_3$ to be $(R_1 - R_3)/R_1=0.087\pm0.017$. Putting this value into context with other USPs, Figure~\ref{fig:tidal}a shows that, under these assumptions, TOI-2431 b is among the most tidally deformed USP planets, along with TOI-6324 b \citep{RenaLee_2025ApJ}, KOI-55 b \citep{2011Natur.480..496C}, and TOI-6255 b \citep{Dai2024-qd}.

% ---------------------------------------------------
\subsubsection{Orbital Decay}
Tidal interactions also result in energy dissipation, leading to a gradual inward spiral of a planet toward its host star. The tidal decay timescale $\tau_P$, assuming a constant-lag-angle model \citep{1966Icar....5..375G, Dai2024-qd}, is given by:
\begin{equation}\label{eqn:tidal_decay}
\tau_P \equiv \frac{P}{\Dot{P}} \approx 30~\mathrm{Gyr} \left(\frac{Q_\star'}{10^6}\right) \left(\frac{M_\star/M_p}{M_\odot/M_\oplus}\right) \left(\frac{\rho_\star}{\rho_\odot}\right)^{5/3} \left(\frac{P}{1~\rm{day}}\right)^{13/3},
\end{equation}
where $Q'_*$ is the reduced stellar tidal quality factor and is the main source of uncertainty for this calculation. Following \cite{Dai2024-qd} and \cite{RenaLee_2025ApJ}, we adopt $Q_\star' = 10^7$, representing a relatively slow stellar dissipation rate as empirically found by \cite{2018AJ....155..165P}. From this, we find $\tau_{P} \approx200 \, \rm{Myr}$ for TOI-2431 b, while stressing the underlying uncertainty in the $Q_\star'$ value of the star that may differ from the assumed value by more than an order of magnitude and would change the timescale accordingly.

The time it takes to reach the Roche limit can be found by integrating Equation~\ref{eqn:tidal_decay}, yielding
\begin{equation}\label{eqn:roche_timescale}
\tau_{Roche} = k\left(\left(\frac{P}{1~\rm{day}}\right)^{13/3} - \left(\frac{P_{Roche}}{1~\rm{day}}\right)^{13/3}\right),
\end{equation}
where 
\begin{equation}
k \approx 7~\mathrm{Gyr} \left(\frac{Q_\star'}{10^6}\right) \left(\frac{M_\star/M_p}{M_\odot/M_\oplus}\right) \left(\frac{\rho_\star}{\rho_\odot}\right)^{5/3},
\end{equation}
resulting in $\tau_{\mathrm{Roche}} \approx31 \, \rm Myr$. We compare $\tau_{\mathrm{Roche}}$ of TOI-2431 b to other well-characterized USPs in Figure~\ref{fig:tidal}b---where we assumed a fixed $Q_\star'=10^7$ for all of the systems---from which we see that TOI-2431 b has the shortest $\tau_{\mathrm{Roche}}$ timescale. This puts TOI-2431 b in a regime where orbital decay could potentially be detectable with long-term observations \citep[see e.g.,][]{2017ApJ...836L..24W}. Such observations will constrain $Q'_*$ and may lead to further insights into the tidal interactions and future evolution of the system.

% -----------------------------------------------------
\subsection{Comparison with other USPs and future observations}\label{sec:comparisonandobservations}
With a period of only 5.4 hours, TOI-2431 b is among the shortest-period exoplanets discovered to date, but is the shortest period planet with both a well-characterized mass and radius. The first and second shortest-period planets, PSR J1719-1438 b \citep[$P=2.2$ h;][]{Bailes2011-vy} and M62H b \citep[$P=3.2$ h;][]{Vleeschower2024-al}, were discovered with the pulsar timing method, which lacks planetary radius measurements. As for the third and fourth shortest-period planets, KOI-1843.03 \citep[$P=4.2$ h;][]{Rappaport2013-ew} and K2-137 b \citep[$P=4.3$ h;][]{Smith2018-hl}, both were discovered by the transit technique, but currently lack robust planet mass measurements. The fifth shortest-period planet, KIC 10001893 b \citep[$P=5.3$ h;][]{Silvotti2014-mh}, has been discovered with the orbital brightness modulation technique and lacks planetary mass and radius measurements.

\begin{figure*}
\centering
\includegraphics[width=\hsize]{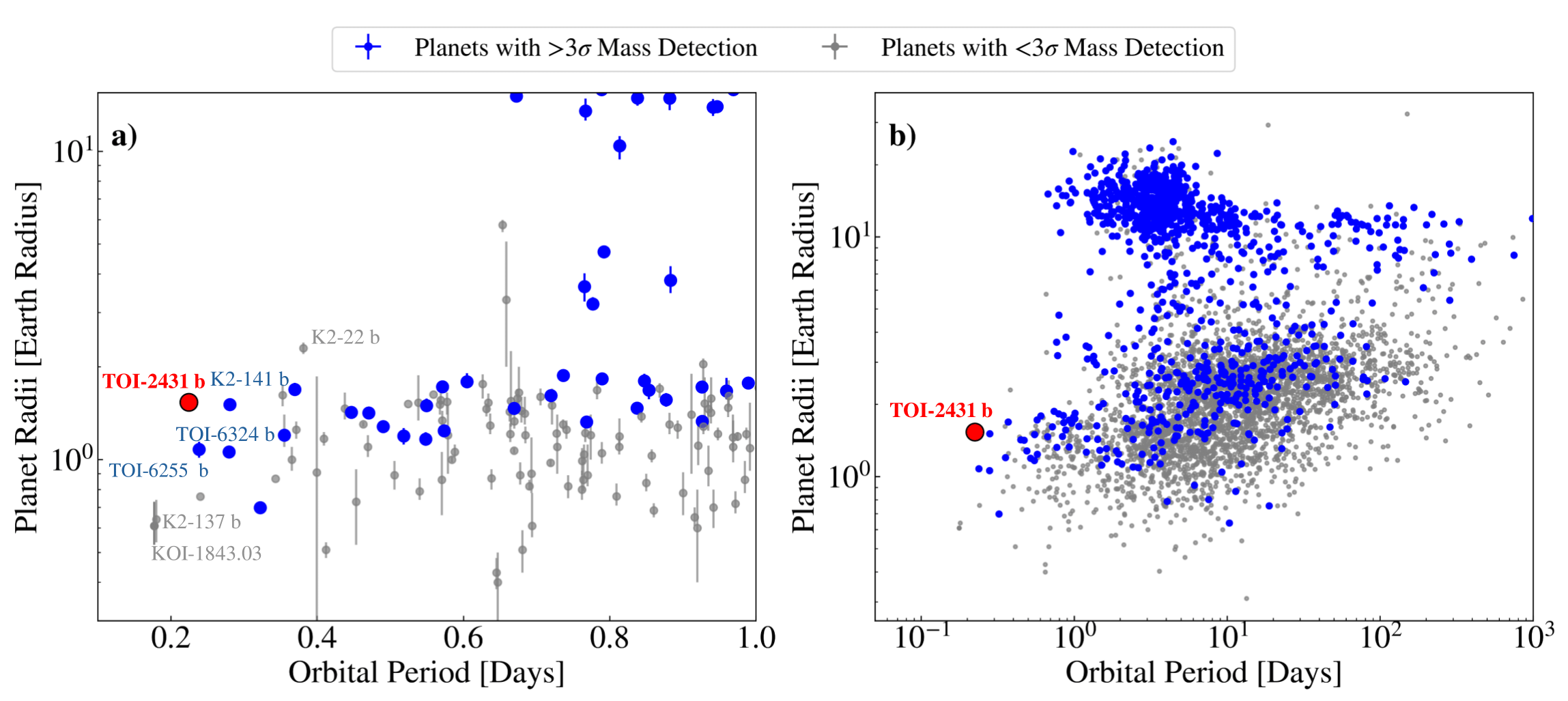} 
\caption{Planet radius as a function of orbital period for known exoplanets. Confirmed planets with a $3\sigma$ or higher precision mass detection are highlighted in blue. The planets that have a mass detection smaller than $3\sigma$ precision are shown in gray. TOI-2431 b is shown in red. TOI-2431 b is the shortest period planet with both a characterized mass and radius (better than $3\sigma$). \textbf{a)} TOI-2431 b compared to other USPs. \textbf{b)} Same as a) but comparing TOI-2431 b to the broader exoplanet population showing that TOI-2431 b is at the lower edge of the Hot Neptune Desert.}
\label{fig:exoplanetpopulation}
\end{figure*}

The closest exoplanets to TOI-2431 b in terms of period and orbital parameters are TOI-6255 b \citep[$P=5.7$ h;][]{Dai2024-qd}, and TOI-6324 b \citep[$P=6.7$ h;][]{RenaLee_2025ApJ}, and K2-137 b \citep[$P=4.3$ h;][]{Smith2018-hl}, which were all first detected using the transit technique. As seen in Figure~\ref{fig:exoplanetpopulation}, TOI-2431 b is one of the USP planets that form the lower boundary of the Hot Neptune Desert, which is a sparsely populated region in the exoplanet population \citep{2016mazehA&A...589A..75M}.

\begin{figure}
\centering
\includegraphics[width=\hsize]{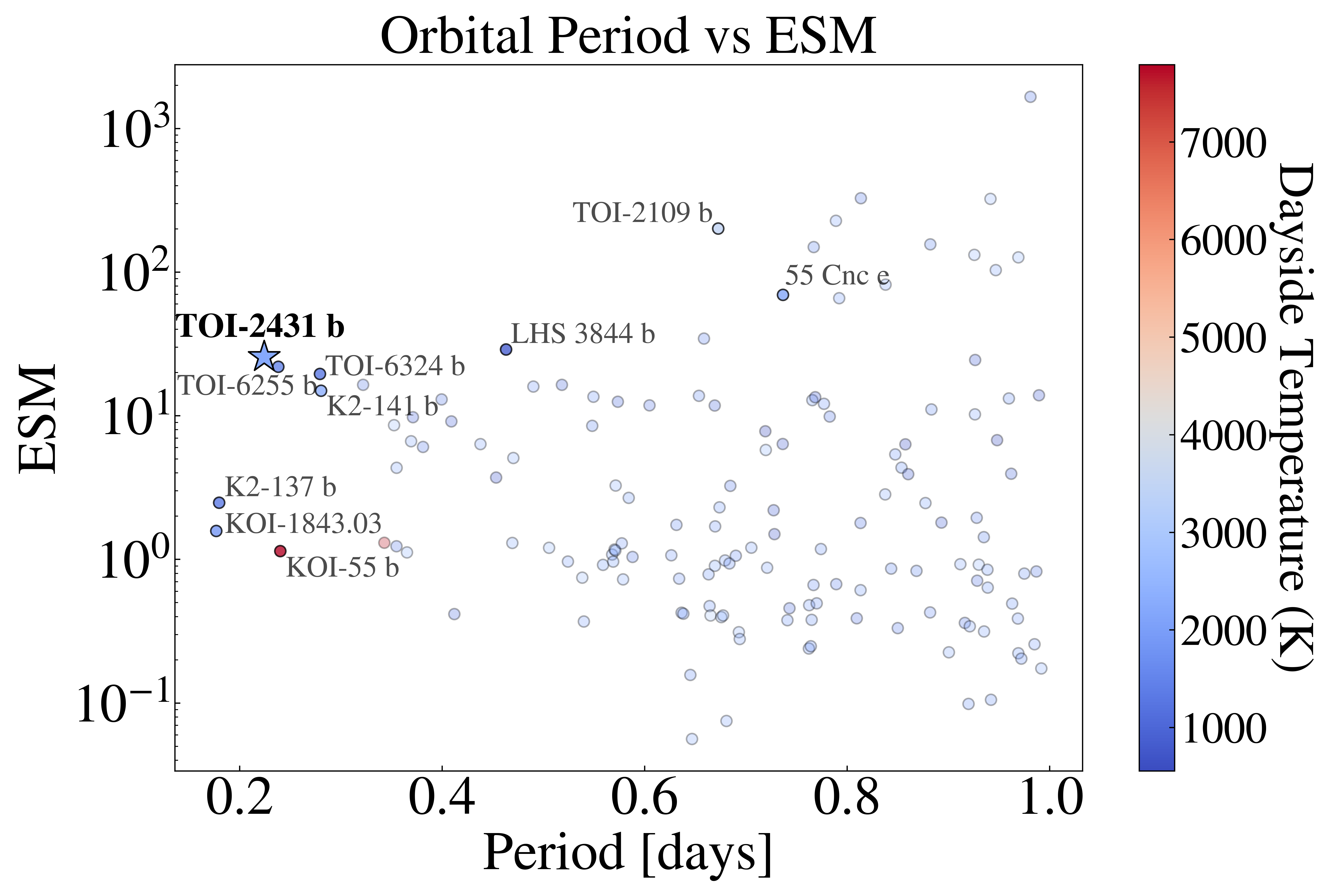}
\caption{JWST Emission Spectroscopy Metric (ESM) \citep{Kempton_2018} as a function of planet orbital period for TOI-2431 b and other confirmed USP planets. TOI-2431 b is highlighted with the star, and other well-characterized USPs are shown with the thicker labels. The color bar shows the dayside equilibrium temperature assuming an albedo of $A_{\rm{B}} = 0$.}
\label{fig:ESM} 
\end{figure}

With its ultra-short period and the brightness of its host star, TOI-2431 b is a promising candidate for phase-curve observations with JWST. Figure~\ref{fig:ESM} shows the Emission Spectroscopy Metric \citep[ESM;][]{Kempton_2018}---which quantifies the feasibility of phase-curve and emission spectroscopic observations with JWST---as a function of orbital period for TOI-2431 b compared to other well-characterized USP planets. TOI-2431 b has an $\mathrm{ESM}\sim27$ (Table~\ref{table:planetaryparams}), which makes it one of the best rocky planets for phase-curve observations. Such JWST phase curve observations, could shed light onto the presence or absence of a heavy-mean-molecular-weight atmosphere on TOI-2431 b \citep{Dai2024-qd}. In the presence of an atmosphere, JWST's ability to detect thermal emission across a broad wavelength range would allow a probe of the temperature distribution difference between the dayside and nightside of the planet, yielding insight into the atmospheric circulation of the planet. In the case of no atmosphere, phase curve observations could inform about the planet's surface mineralogy, yielding valuable insights into its composition and evolutionary history \citep{refId0, 2024_Zhang, 2022_Whittaker}.

% -----------------------------------------------------
% -----------------------------------------------------
% -----------------------------------------------------
\section{Conclusions}\label{sec:conclusion}
We have confirmed the ultra-short period planet TOI-2431 b using a combination of photometric transit data from TESS, precise RV observations with the NEID and HPF spectrographs, and ground-based speckle imaging with the NESSI instrument. With a period of $P=0.224\:\mathrm{days}$, TOI-2431 b is among the shortest period exoplanets detected to date. From a joint analysis the available transit data and RVs, we measure a radius of $R_p =1.536 \pm 0.033 \, \rm{R_\oplus}$, and a mass of $M_p = 6.2 \pm 1.2 \, \rm{M_\oplus}$, suggesting a bulk density of $9.4_{-1.8}^{+1.9}\,\mathrm{g/cm^3}$ which is slightly denser, but consistent with an Earth-like composition. The planet has an equilibrium temperature of $2063\pm 30 \, \mathrm{K}$ (assuming Bond Albedo as $A_{\mathrm{B}}=0$), suggesting the surface is likely molten. We note that we see evidence of excess scatter in our RV data compared to a single Keplerian planet model, which we interpret as suggestive evidence that there are probably other non-transiting planets in the system. We urge additional RV follow-up observations to gain further insights into other planets in the system.

Additionally, we show that TOI-2431 b is close to its Roche limit with $P_{\mathrm{orb}} \approx 1.31 \, P_{\mathrm{Roche}}$. Due to its close-in orbit, we show that TOI-2431 b is likely tidally deformed, with its shortest axis being $\sim$$9\%$ shorter than its longest axis. Furthermore, assuming a nominal tidal quality factor of $Q_\star'=10^7$, we estimate a tidal decay timescale of only $\tau_{\mathrm{Roche}} \approx31 \, \rm Myr$, which is the shortest tidal decay timescale compared to other known USPs. 

Finally, we show that TOI-2431 b has an ESM of 27, making it one of the best USP planets for phase curve observations with JWST, which could shed light into the surface composition, and if the planet has an atmosphere.

% --------------------------------------------------------
% --------------------------------------------------------
% --------------------------------------------------------
\section*{Acknowledgements}
This effort started as an observation proposal for granted observation time at the 1.2m Mercator Telescope as part of the La Palma Observing Class at the University of Amsterdam as a group project led by Master's students Kaya Han Taş, Esha Garg, and Syarief N.M. Fariz. The project grew into the current manuscript under the supervision of Gudmundur Stefansson. Esha and Syarief contributed equally to the manuscript. Kaya, Esha, and Syarief thank Antonija Oklopčić, Rudy Wijnands, Stefanie Fijma, and Nathalie Degenaar for helpful discussions during the observing project.

We acknowledge support from NSF grants AST 1006676, AST 1126413, AST 1310875, AST 1310885, and the NASA Astrobiology Institute (NNA09DA76A) in our pursuit of precision radial velocities in the NIR. We acknowledge support from the Heising-Simons Foundation via grant 2017-0494. This research was conducted in part under NSF grants AST-2108493, AST-2108512, AST-2108569, and AST-2108801 in support of the HPF Guaranteed Time Observations survey. The Hobby-Eberly Telescope is a joint project of the University of Texas at Austin, the Pennsylvania State University, Ludwig-Maximilians-Universitat Munchen, and Georg-August Universitat Gottingen. The HET is named in honor of its principal benefactors, William P. Hobby and Robert E. Eberly. The HET collaboration acknowledges the support and resources from the Texas Advanced Computing Center. We thank the Resident astronomers and Telescope Operators at the HET for the skillful execution of our observations with HPF.

These results are based on observations obtained with the Habitable-zone Planet Finder (HPF) spectrograph on the HET. The HPF team was supported by grants from the National Science
Foundation, the NASA Astrobiology Institute, and the Heising-Simons Foundation.

The Texas Advanced Computing Center (TACC) at the University of Texas at Austin provided high-performance computing, visualization, and storage resources that have contributed to the results reported within this paper.

The Center for Exoplanets and Habitable Worlds and the Penn State Extraterrestrial Intelligence Center are supported by Penn State and its Eberly College of Science.

We thank the NEID Queue Observers and WIYN Observing Associates for their skillful execution of our observations. Data presented were obtained by the NEID spectrograph built by Penn State University and operated at the WIYN Observatory by NOIRLab, under the NN-EXPLORE partnership of the National Aeronautics and Space Administration and the National Science Foundation. The NEID archive is operated by the NASA Exoplanet Science Institute at the California Institute of Technology. 

Based in part on observations at the Kitt Peak National Observatory (Prop. ID 2025A-546977), managed by the Association of Universities for Research in Astronomy (AURA) under a cooperative agreement with the National Science Foundation. The WIYN Observatory is a joint facility of the NSF’s National Optical-Infrared Astronomy Research Laboratory, Indiana University, the University of Wisconsin-Madison, Pennsylvania State University, Purdue University, and Princeton University. The authors are honored to be permitted to conduct astronomical research on Iolkam Du’ag (Kitt Peak), a mountain with particular significance to the Tohono O’odham.

This paper includes data collected by the TESS mission. Funding for the TESS mission is provided by NASA's Science Mission Directorate. This research made use of the NASA Exoplanet Archive and ExoFOP, both of them operated by the California Institute of Technology, under contract with the National Aeronautics and Space Administration under the Exoplanet Exploration Program. This research has made use of the SIMBAD and VIZIER databases at CDS, Strasbourg (France), and of the electronic bibliography maintained by the NASA/ADS system. This work has made use of data from the European Space Agency (ESA) mission Gaia processed by the Gaia Data Processing and Analysis Consortium (DPAC). Funding for the DPAC has been provided by national institutions, in particular the institutions participating in the Gaia Multilateral Agreement.

The research was carried out, in part, at the Jet Propulsion Laboratory, California Institute of Technology, under a contract with the National Aeronautics and Space Administration (80NM0018D0004).

J.I.E.-R. gratefully acknowledges support from the John and A-Lan Reynolds Faculty Research Fund, from ANID BASAL project FB210003, and from ANID Doctorado Nacional grant 2021-21212378. C.I. Cañas acknowledges support by NASA Headquarters through an appointment to the NASA Postdoctoral Program at the Goddard Space Flight Center, administered by ORAU through a contract with NASA.

\bibliographystyle{aa} 
\bibliography{references} 

\appendix

%-------------------------------------------------------------------
%-------------------------------------------------------------------
%-------------------------------------------------------------------
\section{Radial Velocities}\label{sec:rvdata}
Table~\ref{table:rvobservations} lists the RVs of TOI-2431.

\begin{table}[h]             
\centering                          
\caption{RVs of TOI-2431. The median S/N for NEID observations is $27$ at $860 \,\rm{nm}$, and the median S/N of the HPF observations is $205$ at $1\,\mu m$. The RVs used in the FCO fit---the nights that had two or more observations obtained within a night---are denoted with "Y" in the FCO column. All of the RVs were used for the GP joint fit. }\label{table:rvobservations}
\begin{tabular}{lccc}        
\hline\hline                 
 Time ($\text{BJD}_{\mathrm{TDB}}$)& Radial Velocity ($\mathrm{m\,s^{-1}}$) &Instrument &FCO\\ 
\hline                       
 2460661.694342& -9.0 $\pm$ 2.6&NEID &Y\\
 2460661.834318& -21.9 $\pm$ 3.1&NEID &Y\\
 2460663.818467& -1.9 $\pm$ 2.4&NEID &N\\
 2460664.634615& 6.4 $\pm$ 1.8&NEID &N\\
 2460665.583873& 9.8 $\pm$ 1.8&NEID &N\\
 2460680.644771& -12.1 $\pm$ 4.2&NEID &Y\\
 2460680.746552& 6.8 $\pm$ 2.8&NEID &Y\\
 2460685.702258& 4.6 $\pm$ 5.8&NEID &N\\
 2460725.617289& -1.4 $\pm$ 3.0&NEID &Y\\
 2460725.621031& -6.3 $\pm$ 2.7&NEID &Y\\
 2460725.672322& -16.3 $\pm$ 5.6&NEID &Y\\
 2460725.675878& -11.3 $\pm$ 5.4&NEID &Y\\
 2460673.709907& -25.2 $\pm$ 10&HPF &N\\
 2460712.606630& 13.3 $\pm$ 7.3&HPF &N\\
 2460715.597665& 5.9 $\pm$ 5.0&HPF &N\\
\hline
\end{tabular}
\label{tab:rv}
\end{table}

%-------------------------------------------------------------------
%-------------------------------------------------------------------
%-------------------------------------------------------------------
\section{Injection and Recovery Tests for the FCO Method}\label{sec:planetinjections}
To assess the robustness of the Floating Chunk Offset (FCO) method for our joint fitting of the NEID and TESS data in the limit where we had 8 observations spread over 3 NEID visits, we carried out a series of injection and recovery tests.

For the injection and recovery tests, we created a series of synthetic RV datastreams evaluated at the same NEID timestamps as used in the adopted fit in Section~\ref{sec:analysis} and listed in Table~\ref{tab:rv}. The RV uncertainties were assumed to be the same RV uncertainties as the as-observed uncertainties as listed in Table~\ref{tab:rv}. To simulate the impact of the excess RV scatter we see in the system we assumed two planets in the system: planet b, which we assumed has the same orbital period and transit midpoint as our best-fit in Table~\ref{table:planetaryparams}, and another outer planet c in the system. We assumed planet c had an orbital period $P_c$ between 1 to 100 days. For planet c, we randomly sampled the time of periastron from $-P_c/2$ to $P_c/2$. For both planets, we assume circular orbits. We then varied the RV semi-amplitudes of both planets b and c, uniformly sampled from 0 to 50 $\mathrm{m\,s^{-1}}$.

In total, we generated 1500 synthetic such RV datastreams, keeping track of the known injected parameters of planet b and planet c. We then ran these synthetic RV datastreams through our FCO fitting method to test if we accurately recover the parameters of planet b, irrespective of the parameters of the outer companion. Here we define "recovered" if the recovered semi-amplitude of planet b is within $3\sigma$ range of the injected semi-amplitude.

\begin{figure*}[h]
\centering 
\includegraphics[width=\hsize]{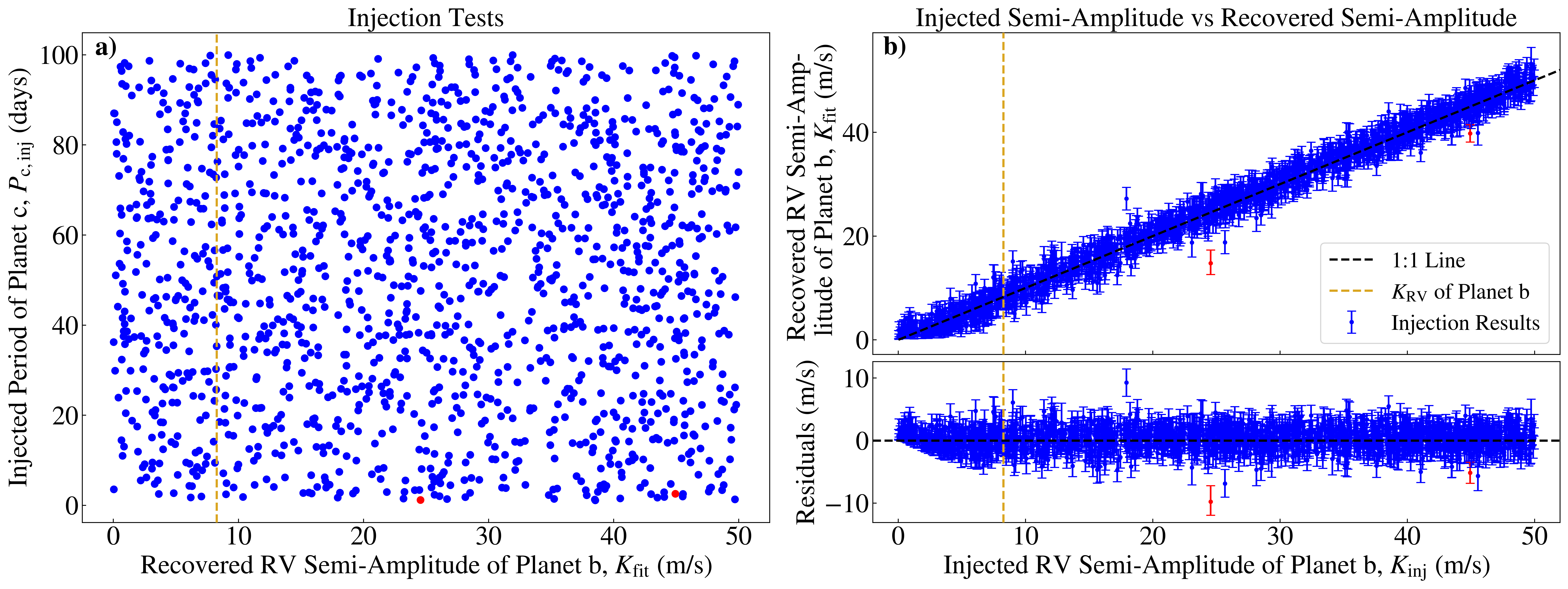} 
\caption{Summary of injection and recovery test to test the robustness of the FCO method in the limit of 8 NEID observations obtained in 3 different visits.  \textbf{a)} Injected semi-amplitudes of a hypothetical planet c as a function of the injected RV amplitude of the assumed planet b. Blue dots represent runs where the RV amplitude of planet b was successfully recovered (within $3\sigma$ of the injected value), while the red dots indicate non-recovered runs. The semi-amplitude of TOI-2431 b from Table~\ref{table:planetaryparams} is highlighted with the yellow vertical line. \textbf{b)} \textit{Top:} Recovered RV semi-amplitudes $K_{\mathrm{fit}}$ as a function of injected RV semi-amplitudes for planet b. Similar to a), blue dots represent recovered runs, whereas red dots indicate non-recovered runs, and the yellow vertical line indicates our adopted RV semi-amplitude of TOI-2431 b from Table~\ref{table:planetaryparams}. The 1-1 line is indicated with the black dashed line. We see that the recovered values are in good agreement with the injected values.}
\label{fig:InjectionComparison}
\end{figure*}

Figure~\ref{fig:InjectionComparison} summarizes the results of the injection and recovery tests. In Figure~\ref{fig:InjectionComparison} blue points show 'recovered' values of planet b, while red points denote cases that were not recovered: out of 1500 injections, we have recovered the semi-amplitude from 1498 of them, suggesting a 99.87\% recovery rate. This frequency corresponds to the $3\sigma \sim 99.7\%$ definition of recovered from the total population. 

Figure~\ref{fig:InjectionComparison} also highlights the recovered semi-amplitudes as a function injected RV semi-amplitudes. This shows that there is no bias introduced in the recovered RV semi-amplitudes of the synthetic planet b at the RV amplitude of TOI-2431 b of 8.3 $\mathrm{m\,s^{-1}}$. From this, we conclude that the FCO method, in combination with the available NEID RV observations results in a robust determination of the RV amplitude of TOI-2431 b.

\end{document}